\documentclass{aa} 

\usepackage{graphicx}
\usepackage{txfonts}
\usepackage{hyperref}
\usepackage{graphicx}	
\usepackage{amsmath}	
\usepackage{amssymb}	
\usepackage{multirow}
\usepackage{booktabs}
\usepackage{textcomp}
\usepackage{gensymb}
\usepackage{siunitx}
\usepackage{subcaption}
\usepackage{epstopdf}
\usepackage{multirow}
\usepackage{caption} 
\usepackage{xcolor}
\usepackage{soul}
\usepackage{anyfontsize}
\usepackage{threeparttable}
\usepackage{array}
\usepackage{multicol}
\usepackage{longtable}

\usepackage{etoolbox} 

\newcommand*\linenomathpatch[1]{%
  \expandafter\pretocmd\csname #1\endcsname {\linenomath}{}{}%
  \expandafter\pretocmd\csname #1*\endcsname{\linenomath}{}{}%
  \expandafter\apptocmd\csname end#1\endcsname {\endlinenomath}{}{}%
  \expandafter\apptocmd\csname end#1*\endcsname{\endlinenomath}{}{}%
}

\newcommand*\linenomathpatchAMS[1]{%
  \expandafter\pretocmd\csname #1\endcsname {\linenomathAMS}{}{}%
  \expandafter\pretocmd\csname #1*\endcsname{\linenomathAMS}{}{}%
  \expandafter\apptocmd\csname end#1\endcsname {\endlinenomath}{}{}%
  \expandafter\apptocmd\csname end#1*\endcsname{\endlinenomath}{}{}%
}

\newcommand\rs[1]{_\mathrm{_{#1}}}
\newcommand\ms[1]{\mathrm{#1}}

\begin{document}

   \title{Optical and X-ray timing analysis of the 2018-2020 outburst and rebrightening of the black-hole transient MAXI J1820+070}

    \author{M. Fiori,$^{1}$\fnmsep\thanks{E-mail (MF): michele.fiori@inaf.it}
            L. Zampieri,$^{1}$\fnmsep\thanks{E-mail (LZ): luca.zampieri@inaf.it}
            A. Burtovoi,$^{2}$
            G. Naletto,$^{3,1}$
            P. Ochner,$^{1,3}$
            U. Munari,$^{4}$
            F. Manzini,$^{5}$
            A. Vagnozzi,$^{5}$
            E. A. Barsukova,$^{6}$
            M. A. Burlak,$^{7}$
            V. P. Goranskij,$^{7,6}$
            N. P. Ikonnikova,$^{7}$
            N. A. Katysheva,$^{7}$
            E. G. Sheyanov,$^{8}$
            S. Yu. Shugarov,$^{7,9}$
            A. V. Zharova,$^{7}$
            A. M. Zubareva$^{8,7}$
            and S. E. Motta$^{10}$\\}

   \institute{$^{1}$INAF-Osservatorio Astronomico di Padova, Vicolo dell'Osservatorio 5, I-35122, Padova, Italy\\
              $^{2}$INAF - Osservatorio Astrofisico di Arcetri, Largo E. Fermi 5, I-50125 Firenze, Italy \\
              $^{3}$Department of Physics and Astronomy, University of Padova, Via F. Marzolo 8, I-35131, Padova, Italy\\
              $^{4}$INAF-Osservatorio Astronomico di Padova, Via dell'Osservatorio 8, I-36012, Asiago (VI), Italy\\
              $^{5}$ANS Collaboration, c/o Astronomical Observatory, I-36012, Asiago (VI), Italy\\
              $^{6}$Special Astrophysical Observatory, Russian Academy of Sciences, Nizhnij Arkhyz, Karachai-Cherkesia, 369167, Russia\\
              $^{7}$Sternberg Astrophysical Institute, Moscow University, Universitetsky Ave., 13, Moscow 119992, Russia\\
              $^{8}$Institute of Astronomy (Russian Academy of Sciences), Pyatnitskaya Str. 48, 119017 Moscow, Russia\\
              $^{9}$Astronomical Institute of the Slovak Academy of Sciences, Tatransk\'a Lomnica, 059 60, The Slovak Republic\\
              $^{10}$INAF-Osservatorio Astronomico di Brera, via E. Bianchi 46, I-23807, Merate (LC), Italy}

   \date{Received ...; accepted ...}

  \abstract{We report the results of a comprehensive analysis of the multiwavelength (in optical and X-rays) and multitimescale (from months to tenths of a second) variability of the 2018-2020 outburst of the black hole transient MAXI J1820+070. During the first outburst episode, a detailed analysis of the optical photometry shows a periodicity that evolves over time and stabilises at a frequency of $1.4517(1)$ 1/d ($\sim0.5$\% longer than the orbital period). This super-orbital modulation is also seen in the X-rays for a few days soon after the transition to the high-soft state. We also observed optical Quasi-Periodic Oscillations (QPOs), which correspond to some of the QPOs observed in X-rays at three different epochs when the source was in the low-hard state. In two epochs, optical QPOs with a centroid consistent with half the frequency of the most prominent X-ray QPO can be seen. If the lowest modulation frequency is the one observed in the optical, the characteristic precession frequency of MAXI J1820+070 is lower than that inferred from the `fundamental' QPO in the X-rays. Assuming that QPOs can be generated by Lense-Thirring precession, we calculate the spin of the black hole in the case where the fundamental precession frequency is tracked by the optical emission. We find a relatively slowly spinning black hole with a spin parameter {$\lesssim 0.15$}. The super-orbital optical and X-ray modulations observed after the disappearance of the QPOs may be triggered by the self-irradiation of the outer disc by a standard inner disc truncated at a few gravitational radii.}


    \titlerunning{Optical and X-ray timing analysis of MAXI J1820+070}
    \authorrunning{Fiori et al.}
   \maketitle
%


\section{Introduction}

A large fraction of the transient sources can be associated with systems containing compact objects. 
Being able to study the emission of these systems simultaneously at multiple wavelengths and on multiple time scales has proven to be of fundamental importance in helping us to correctly interpret physical phenomena that are still poorly understood and to be able to estimate the fundamental properties of the compact objects that power them. 

MAXI J1820+070 is a bright X-ray black hole transient discovered on 06 March 2018 in the optical band (ASASSN-18ey; \citealt{ATel11400,Tucker2018}) by the All-Sky Automated Survey for SuperNovae (ASAS-SN; \citealt{asassn2014}) and six days later in the X-ray band \citep{ATel11399} by the Monitor of All-sky X-ray Image (MAXI; \citealt{MAXI2009})\footnote{Other common names for this star are V3721 Oph \citep{Kazarovets2019II}, Gaia18asi, WISE 182021.94+071107.2. A more complete list of designations and a summary of the main characteristics of the object can be found at \url{https://www.aavso.org/vsx/}}. The main outburst lasted several months, roughly until the end of 2018, and has been followed by a series of subsequent rebrightenings that were recorded until June 2020 (plus some weak activity reported in March and April 2021; \citealt{ATel14492, ATel14582}).
During this long outburst, the source was widely observed at different wavelengths, from the radio band up to the X-rays. No detection was reported at $\gamma$-ray energies \citep{GammaRayMAXIJ1820_2022}.

The high luminosity attained by MAXI J1820+070, due in part to the relatively small distance of the source \citep[$d=2.96\pm0.33$ kpc;][]{Atri2020}, and the complex phenomenology displayed during the outburst made this source a target for many observational campaigns at different wavelengths \citep{Kalemci2022}.
In the first months of the outburst, the source went through all the typical states and transitions of a black hole accreting X-ray binary (Low/Hard state: LH;  High/Soft state: HS; Intermediate state: IM), as reported in detail by \cite{Shidatsu19}. For a review on the different accretion states of X-ray binaries see e.g. \cite{Remillard2006} and \cite{Done07}. 

The mass of the black hole was estimated from the spectral shifts and the ellipsoidal modulation of the optical emission of the companion star during quiescence ($\ms{M}\rs{BH} = 8.48^{+0.79}_{-0.72}$ $\ms{M}_{\odot}$; \citealt{Torres2020}). Some of the jet properties, such as the inclination angle ($\theta\rs{JET} = 63 \pm 3$ deg) and the velocity ($\ms{v} = 0.89 \pm 0.09$ c), have been estimated from European VLBI radio measurements \citep{Atri2020}. Other estimates of the jet properties derived from modelling the highly correlated emission between different bands confirmed the existence of a high relativistic and confined jet ($\Gamma = 6.81^{+1.06}_{-1.15}$, $\phi = 0.45^{+0.13}_{-0.11}$ deg) that carries away a good fraction of the total power \citep[$\sim 0.6$ $\ms{L}\rs{1-100keV}$;][]{Tetarenko2021}.

The properties of these systems can also be studied using fast timing features observed in the Fourier domain. The Power Density Spectra (PDS) often reveal broad and narrow features related to particular phenomena occurring in the accretion flow. Variability in narrow frequency ranges are usually referred to as Quasi-Periodic Oscillations (QPOs). QPOs can be divided into Low-Frequency QPOs (LFQPOs, $\sim0.005-40$ Hz) and High-Frequency QPOs (HFQPOs, $\sim40-450$ Hz). The former are further divided into different types based on their properties (type-A, type-B and type-C QPOs). For a description of the different types of QPOs see for example \citep{IngramMotta2019} and reference therein. 
When MAXI J1820+070 was in the LH state a type-C LFQPO with characteristic frequencies below 1 Hz was detected in the X-ray band with Swift/XRT and NICER \citep{Stiele20}. 
Around the time of the first state transition (from the LH to the HS state) the LFQPO switched quickly from type-C to type-B together with the emission of a strong radio flare. These facts were interpreted in terms of the launch of a superluminal jet \citep{Bright2020, Homan2020}.
Broad-band noise features in the PDS of MAXI J1820+070 were also reported and interpreted again in terms of a truncated accretion disk \citep{Dzielak2021,Yang2022}.

LFQPOs synchronous to the X-ray QPOs were also observed at optical wavelengths with various instruments \citep{Zampieri2019, Mao2022, ThomasQPOs2022}. A time delay of $\sim165$ ms was found cross-correlating the X-ray and optical light curves, with the optical lagging the X-ray \citep{Paice2019}. These findings are compatible with the hypothesis that the emission in the two bands is produced in close by regions, in the inner accretion flow and/or at the base of a precessing jet \citep{Paice2019, ThomasQPOs2022}. In a later work, analysing further simultaneous optical/X observations \cite{Paice2021} found that the emission can be separated into two distinct synchrotron components originating from a compact jet and a hot flow, respectively.

Looking at the variability at longer timescales, \cite{patterson18}, reported a $\sim$16.87 hr optical modulation appearing some days before the transition to the HS state (corrected to $\sim$16.57 hr in a later work, \citealt{Patterson19}). This modulation was interpreted as a super-orbital motion (superhump) emission since it differs from the orbital ones ($\sim$16.45 hr, \citealt{Torres2019}). Superhumps are often observed in cataclysmic variables (CVs) and are large-amplitude photometric oscillations with a period that is longer then the orbital period of the binary system (the phenomenon of superhumps in CVs is explained in detail, for example, in \citealt{Kato2009, Kato2017}). 
Also in \cite{Niijima2021}, where a detailed photometric study of MAXI J1820+070 is presented, these oscillations are interpreted as a process similar to the superhumps seen in many CVs.
\cite{ThomasSupehump2021} proposed that the large optical modulation in MAXI J1820+070 could be caused by a warped precessing accretion disc \citep{Ogilvie2001}.

In principle, the quality of the data available for MAXI J1820+070 allows us to perform rather accurate estimates of the spin of the black hole. However, diverse methods bring us to quite different results.
Using the continuum-fitting method with a thin disk model \citep{Zhang1997} and the X-ray soft-state spectra taken with \textit{Insight}-HXMT \citep[Hard X-ray Modulation Telescope;][]{Zhang2020}, \cite{Guan2021} and \cite{Zhao2021} found a relatively slow spinning black hole, with spin parameter a$_*$ equal to $0.2^{+0.2}_{-0.3}$ and $0.14\pm0.09$ respectively. On the other hand, using some characteristic frequencies in the PDS \citep{Motta2014} and the Relativistic Precession Model \citep[RPM][]{Stella1998}, \cite{Bhargava2021} estimated a fast spinning black hole with a*$=0.799^{+0.016}_{-0.015}$. 
Further uncertainties on the validity of the measurements of the spin of the black hole in MAXI J1820+070 come also from optical polarimetric observations, that led to place a lower-limit of 40° on the angle between the spin and orbital axes \citep{Poutanen2022}. As reported by \citep{Poutanen2022} 
\\

The purpose of this work is to try to provide the most comprehensive view of the variability in two different bands (optical and X-ray) and on different time scales of the 2018-2020 outburst of MAXI J1820+070. To this end, we made use of public and proprietary optical and X-ray data from various telescopes and resources, with different time resolutions.

The paper is structured as follows. In section \ref{sec:observations} we briefly describe the observations and the data reduction process. In section \ref{sec:analysis} we show the analysis made on the data, starting from the analysis of the low cadence data, moving then to the sub-second time-scale and finally to the spectroscopic data. In section \ref{sec:discussion} we discuss our results and in section \ref{sec:conclusions} we report our conclusions.

\section{Observations and data reduction}\label{sec:observations}
 For the low cadence photometric data, we used observations taken with the Schmidt telescope in Asiago, the telescopes of the ANS Collaboration\footnote{\url{http://www.ans-collaboration.org/}} \citep{ANScoll2012}, and observations from the AAVSO\footnote{\url{https://www.aavso.org/}} (American Association of Variable Stars Observers, \citealt{AAVSO}). For the subsecond optical variability, we used our own observations collected with the fast photon counters IFI+Iqueye \citep{IQUEYE} and Aqueye+ \citep{AQUEYE} in Asiago . Finally, for the X-ray light curve and fast variability, we exploited the rich dataset collected with the NICER satellite \citep{NICER2012, NICER16}. We also used a series of low resolution spectroscopic data taken with the Asiago 1.22m telescope. The following subsections report a summary of the entire optical and X-ray dataset used in this work.

\subsection{Optical photometry}
In Table \ref{tab:log_lowtime} we have summarized the log of the photometric observations of MAXI J1820+070. The observations span almost 2.5 years (from March 2018 to August 2020) and the dataset comprises $\sim2\times10^5$ photometric measurements.
We have obtained the optical photometry of MAXI 1820+070 in the UBVRI system defined by the equatorial standards of \cite{Landolt1992} and in the $g'$$r'$$i'$ Sloan bands as given in the APASS All-Sky Survey \citep{Henden2014, Henden2018}. The observations come from multiple facilities in Italy, Slovakia, Crimea and Russia.  The telescopes involved in the observations are reported in Tab. \ref{tab:telescopes}.

Data reduction has involved all the usual steps for bias, dark and flat, with calibration images collected during the same observing nights. We adopted aperture photometry because the sparse field around MAXI 1820+070 did not require PSF-fitting procedures.  The transformation from the local to the standard system was carried out via nightly colour equations calibrated on a photometric sequence recorded on the same frames and extracted from the APASS survey, ported to the Landolt system via the transformations calibrated by \cite{Munari2014}.  
The final errors are the quadratic sum of the Poissonian error on the variable and the error in the transformation to the standard system via the instantaneous colour equations. Typical errors are of the order of few hundredths of magnitudes. 

We included in the analysis a part of the 2018 data from the AAVSO \citep{AAVSO}. These data were used to complement those taken with our instrumentation, especially for filling periods when the sampling was scanty. These data are already reduced and calibrated and can be freely downloaded from the AAVSO website.\\

Using the entire cleaned and calibrated dataset, we searched the periodicities in the light curve of MAXI J1820+070 as described in section \ref{sec:opt_phot}. 

\subsection{Optical spectroscopy}

Low resolution spectroscopy of MAXI J1820+070 was obtained with the 1.22m telescope + B\&C spectrograph operated in Asiago by the Department of Physics \& Astronomy of the University of Padova.  The CCD camera is a ANDOR iDus DU440A with a back-illuminated E2V $42-10$  sensor, $2048\times512$ array of 13.5 $\mu$m pixels.  It is highly efficient in the blue down to the atmospheric cut-off around 3200 \AA, and it is normally not used longward of 8000 \AA $ $ for the fringing affecting the sensor.  The adopted 300 ln/mm grating, blazed at  5000 \AA, allowed to cover the wavelength range from $\sim3200$ to $\sim8000$ \AA $ $ at a spectral dispersion of 2.31 \AA/pix.  The slit  width was set to 2-arcsec, providing a resolution of FWHM(PSF)=$2.2$ pix. The slit was always aligned with the parallactic angle for optimal absolute flux calibration, which was achieved by observations of the spectro-photometric standard HR 6900.  This standard is conveniently located just a few degrees away from MAXI J1820+070, and has a similarly hot energy distribution and blue colours, to the benefit of a higher quality of the flux calibration.  BVR magnitudes were computed by band-profile integration on all recorded spectra, and were checked for consistency against nearly simultaneous CCD photometry collected with the aim of building the light and colour-curves of MAXI J1820+070 described in this paper.  MAXI J1820+070 was observed at 10 epochs distributed through the 1st to the 5th maxima of the optical light curve, each time for about 2700 s and the recorded spectra are presented in Fig. \ref{fig:spectra_all} were observing dates are provided. 

\subsection{High Timing Resolution Optical Observations}

We performed 8 observational runs of MAXI J1820+070 with the IFI+Iqueye (IQ) and Aqueye+ (AQ+) fast photon counters \citep{Barbieri2009,IQUEYE,AQUEYE} mounted at the 1.2 m Galileo telescope and the 1.8 m Copernicus telescope (Asiago, Italy) from April to October 2018 (summary of observations in Table \ref{tab:log_hightimeres}). The source was observed in white light (without filter). The data were reduced with the QUEST software (v. 1.1.5, see \citealt{AQUEYE}).
Since the instruments can only observe one source at a time (having a few arcseconds field of view), in between on-target acquisitions, we also regularly observe the sky and a nearby star. The selected star, GSC 00444-02282, was observed for calibration purposes (to be able to properly compare data taken on different nights) and as a reference to check for possible systematics in the power density spectra (PDS) of MAXI J1820+070. The data taken on source during a night are divided into segments of $\sim30/60$ minutes each, while a typical observation of the reference star is of the order of 15 minutes.

The final output of the low-level analysis consists of a sequence of time series that are subsequently time-binned to produce the light curves that can be used for the analysis that follows. The selected value of the bin size is 1 ms.

\subsection{X-ray observations}

During 2018, NICER, actively monitored MAXI J1820+070 via the X-ray timing instrument \citep[XTI;][]{NICER2012},  with almost one observation per day between 12 March and 21 November (210 observations in 254 days). We reduced the data between 12 March and 15 October 2018 (ObsID 1200120101-1200120278).  Observations usually have exposure times greater than 1 ks (one observation reaches $\sim22$ ks) for a total exposure of $\gtrsim400$ ks.
We applied the standard data processing procedure\footnote{\url{https://heasarc.gsfc.nasa.gov/docs/nicer/analysis_threads/}}, using the script \texttt{nicerl2} - part of \texttt{HEASoft} (\texttt{v.6.29}) software\footnote{\url{https://heasarc.gsfc.nasa.gov/docs/software/heasoft/}} -- with version 20210707 of the calibration files.
The selected energy range for the analysis is $1-12$ keV and the chosen time bin is 1 ms. We finally barycentred the data with the script \texttt{barycorr}.

X-ray data were used for comparison with optical results, both for the photometry data (rebinning the X-ray light curve with longer time bins) and for the high time resolution data.

\section{Data analysis and results}\label{sec:analysis}

\subsection{Analysis of the optical photometry and the X-ray light curve}
\label{sec:opt_phot}

\subsubsection{Optical photometry}
\begin{figure*}
    \centering
    \includegraphics[width=1.\textwidth]{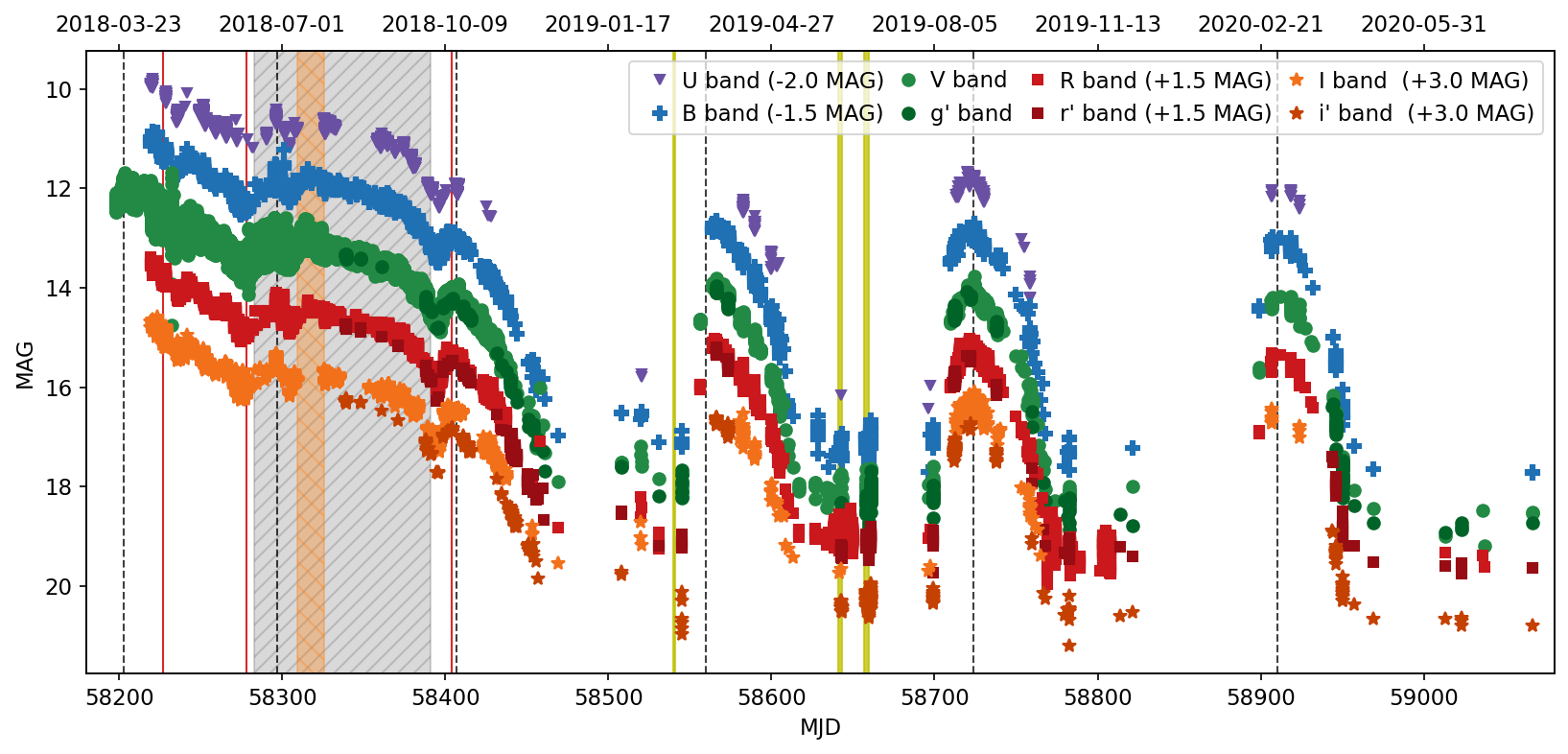}
    \caption{Light curves of MAXI J1820+070 in different optical bands since the epoch of the first maximum after the discovery of the source, and until August 2020. The data taken in the Landolt UBVRI bands \citep{Landolt1992} are shown in violet, blue, light green, light red, and light orange, respectively. The darker green, red, and orange represents the \textit{g'}, \textit{r'}, and \textit{i'} Sloan bands of the APASS system \citep{Henden2014, Henden2018}. The black vertical dashed lines indicate approximately the dates when the local maximum of the optical luminosity is reached during all the subsequent bursts. The red vertical lines indicate the epochs when we {observed} optical LFQPOs with Iqueye. The yellow vertical lines indicate the dates when \citet{Torres2019} measured the periodicity of the binary system ($1.4591$ 1/d)  spectroscopically. The gray shaded area indicates the interval in which a photometric periodicity is detected in this dataset, while the orange shaded area indicates the interval in which a periodicity can be seen also in the X-ray data.}
    \label{fig:lc_longtime_not_corrected}
\end{figure*}

Figure \ref{fig:lc_longtime_not_corrected} shows the optical light curve of MAXI J1820+070 in all the available bands. The data start around the time of the first maximum and extend for more than 2 years. 5 subsequent rebrightenings can be seen after the main burst (the vertical dashed lines show approximately the time of the maximum for each burst).
The intervals between the maxima of all bursting events are: 94 days between the first and second; 110 days between the second and third; 153 days between the third and fourth; and finally 186 days between the last two bursts. The bursting activity is observed in all bands with similar trends. Interestingly, the intervals between two subsequent bursts increase progressively.

In order to search for possible periodicities in the data, we first remove the flaring activity from the long-term light curve {(which is on timescales much longer than the searched periodicities)}. The whole procedure was carried out using a single filter, to avoid possible systematics deriving from averaging together different bands. We relied on V-band data, as they have the best temporal coverage.  We fitted a spline to the average magnitude computed on $N$ consecutive days. After subtracting the spline (detrending), we used an implementation of the Generalized Lomb-Scargle algorithm \citep[LS;][]{lomb1976, scargle1982, lombscargle2017} from the \texttt{Astropy} library \citep{astropy2018} to finally search for periodicities in the data (see Figure \ref{fig:LS_Vband}). {It is worth nothing that a periodicity appears in the periodogram even before applying the detrending procedure, with several spurious peaks caused by the long trend flaring activity.}

\begin{figure}
    \centering
    \includegraphics[width=1.\columnwidth]{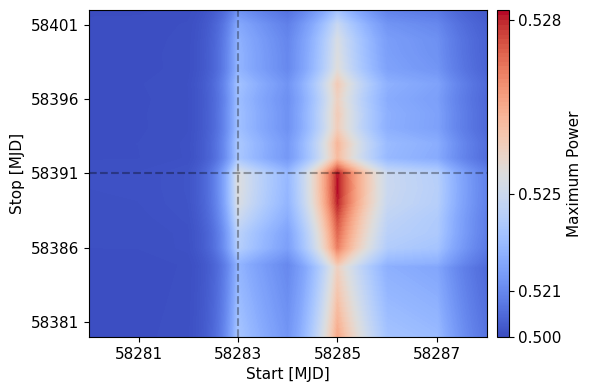}
    \caption{Diagram showing the power of the highest peak in the LS diagram of the V-band light curve after detrending the data and considering different starting and ending dates. This plot was computed to understand which temporal interval to use for searching the best timing solution of the system by means of fitting the sinusoid from equation \eqref{eq:sinusoid} to the data. The power increases from blue to red. A vertical straight line marks the time (MJD 58283) at which the power starts to rise, while a horizontal straight line marks the time (MJD 58391) at which the power starts to decrease.}
    \label{fig:pow_interval}
\end{figure}

\begin{figure}
    \centering
    \includegraphics[width=1.\columnwidth]{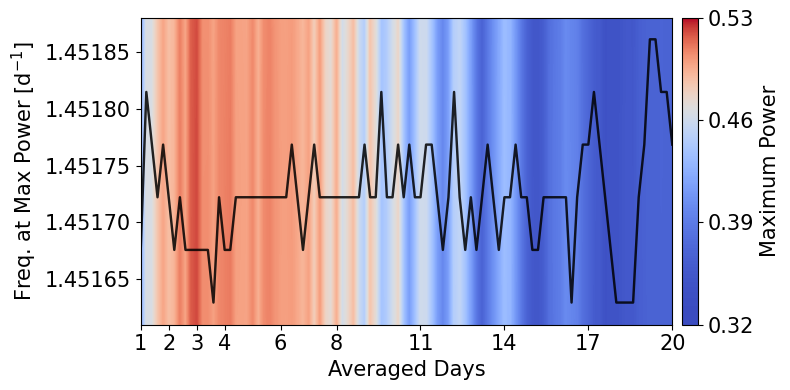}
    \caption{Frequency (black line) and Power (background colour chart) of the highest peak in the LS diagram computed after detrending the V-band data using the average over $N$ days ($x$-axis). The maximum power is reached for $N=3$ days. The spread of the frequency is compatible with the uncertainties of the measurement.}
    \label{fig:freq_pow}
\end{figure}

\begin{figure}
    \centering
    \includegraphics[width=1.\columnwidth]{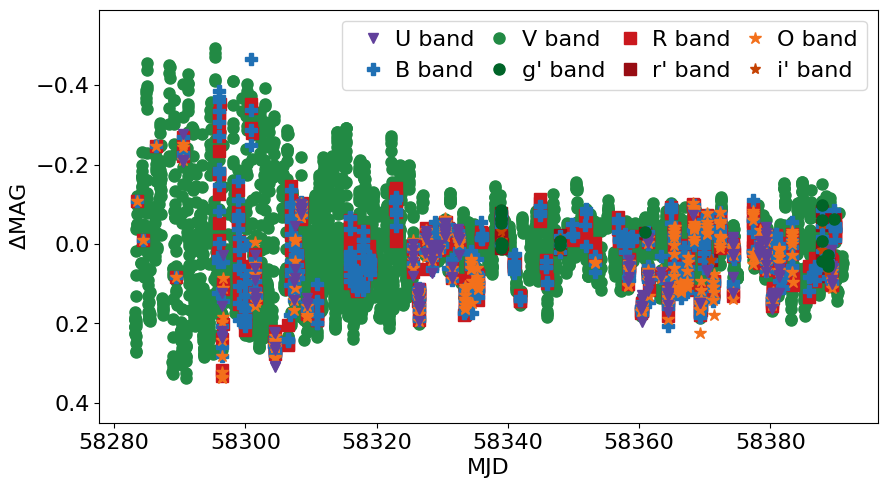}
    \caption{Detrended optical data in the time interval selected to measure the periodicity. The colour code is the same as in Figure \ref{fig:lc_longtime_not_corrected}.}
    \label{fig:cleaned_data}
\end{figure}

\begin{figure}
    \centering
    \includegraphics[width=1.\columnwidth]{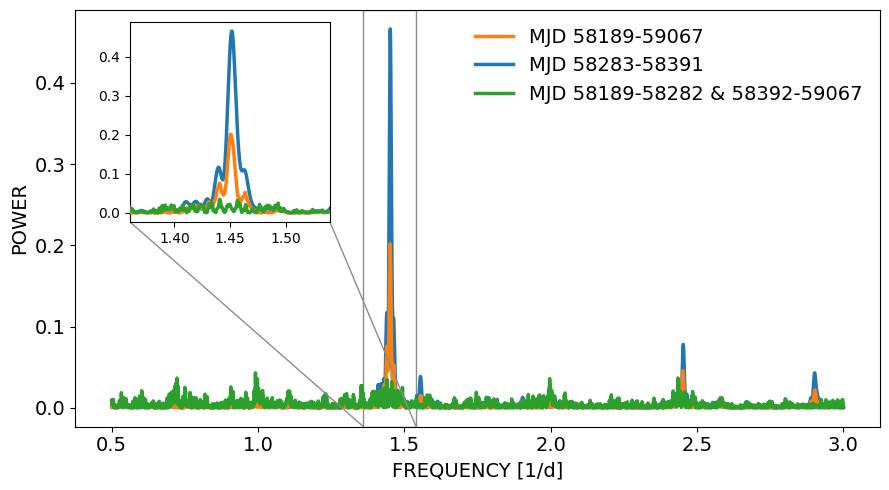}
    \caption{Comparison of LS diagrams in different time intervals. The LS diagram computed considering all the available V-band data is shown in orange, whereas that computed considering only the data within/outside the interval MJD 58283-58391 is shown in blue/green.}
    \label{fig:LS_Vband}
\end{figure}

We initially searched for periodicities on the entire time interval using $N=$1, 2, 3, or 4 days to compute the average magnitude. Results consistently showed a clear peak at a frequency around 1.45 $d^{-1}$. However, folding the whole dataset with this frequency gives a very noisy profile. We therefore tried to restrict the analysis to different temporal sub-intervals to find out when the most prominent periodic signal is generated. We checked that the detrending procedure did not introduce spurious artifacts in the light curve and searched for the interval with the best signal. In the following we describe the adopted iterative procedure:
\begin{enumerate}
    \item We first {clean} the data computing the average over 2 days.
    \item We then searched for a periodicity, computing the LS diagram and looking for the most prominent peak in time windows of 100 days, from the first available date (MJD 58189) to the end (MJD 59066). We found that a peak in the LS diagram appears when we consider a time window between MJD 58285 and MJD 58385.
    \item We computed again the LS diagram changing the starting date in steps of 1 day in an interval of {20 days} around MJD 58285. We did the same also for the ending date. {In this way, the size of all time windows is approximately the same, and we can then compare the maximum powers (next point).}
    \item Figure \ref{fig:pow_interval} shows the maximum power of the peak for different starting and ending dates. We selected the most favourable interval considering as lower bound the date when the power starts rising and as upper bound the date when it starts decreasing. Before MJD 58283 the power is somehow constant (it varies very little) and after that date it starts rising quickly, reaching a maximum at MJD 58285. After MJD 58391 the power begins to decrease slowly and monotonically. We then selected the interval MJD 58283-58391.\footnote{We have chosen MJD 58283 as the starting date rather than MJD 58285 because the power does not decrease between these two dates. On the other hand, before MJD 58283 the power starts to monotonically decrease, as it does after the chosen end date (MJD 58391). In any case, the choice between MJD 58283 and MJD 58285 is not crucial because selecting one date or the other does not have a big impact on the final value of the frequency (see also the top panel in Fig. \ref{fig:diff_mjd_start}).}
    \item After selecting the most favourable interval we returned to step (1) and varied the number of days to average together the data. We detrended the data using different number of days, from 1 to 20 days. With the different detrended data we computed the LS diagram in the interval MJD 58283-58391 and determined the frequency and the power of the most prominent peak (Figure \ref{fig:freq_pow}). The black line indicates the frequency at which we found the maximum power in the LS after detrending with $N$ averaged days ($x$-axis). The background colour chart corresponds to the value of the power at the peak. We found that peak power is maximized averaging together 3 days of data.
    \item After fixing $N=3$ days, we repeated steps (3) and (4) to verify the choice of the interval (confirming that the best interval seems to be between MJD 58283 and MJD 58391).
    \item Finally, the data were rebinned in intervals of 20 minutes to compensate for the different exposure times in the original data-set and to filter out the noise of the light curve at shorter time scales.
\end{enumerate}
The steps in the procedure outlined above may in principle introduce an additional uncertainty in the estimate of the periodicity. However, this does not seem to be a serious concern as in the fitting procedure reported below the error of the fitted frequency turns out to be of the same order of the dispersion visible in Figure \ref{fig:freq_pow}.
Figure \ref{fig:cleaned_data} shows the detrended V-band data in the selected interval, which corresponds to the gray shaded area in Figure \ref{fig:lc_longtime_not_corrected}. Even after the detrending procedure is applied, the curve shows a quite pronounced intrinsic dispersion that varies inside the interval. The amplitude of the variability is large at the beginning, while in the second half of the interval is smaller. The LS diagram in Figure \ref{fig:LS_Vband} shows a strong peak at a frequency of $1.4517\; d^{-1}$, for both all the V-band data (orange curve) and the sole data between MJD 58283-58391 (blue curve). The peak for the data in the selected interval is much higher, while that for the data outside this interval (green curve) completely disappears{, meaning that the periodicity outside this interval is completely absent}. \newline

In order to accurately determine the modulation period, we performed a fit of the detrended light curve considering the following sinusoidal function ($\Delta(t)$):
\begin{equation}\label{eq:sinusoid}
        \Delta(t) = A\sin(2 \pi f [t-t_0] + \phi) + C,
    \end{equation}
where $A$ is the amplitude, $f$ the frequency, $\phi$ the phase with respect to the initial time $t_0$ and $C$ is an offset (to account for possible residuals in the detrending procedure).
To fit the data to equation (\ref{eq:sinusoid}) and estimate the uncertainties of the fitted parameters, we implemented a Markov Chain Monte Carlo procedure (MCMC) through the \texttt{python} package \texttt{emcee} \citep{emcee2013}.
The results of the fitting procedure are reported in Table \ref{tab:fit_results}. The value of the frequency is in agreement with that measured in the LS diagram computed in the same interval. The error is also compatible with the dispersion of the frequency for different choices for the number of nights used in the detrending procedure (see Figure \ref{fig:freq_pow}).

\begin{figure}
    \centering
    \includegraphics[width=1.\columnwidth]{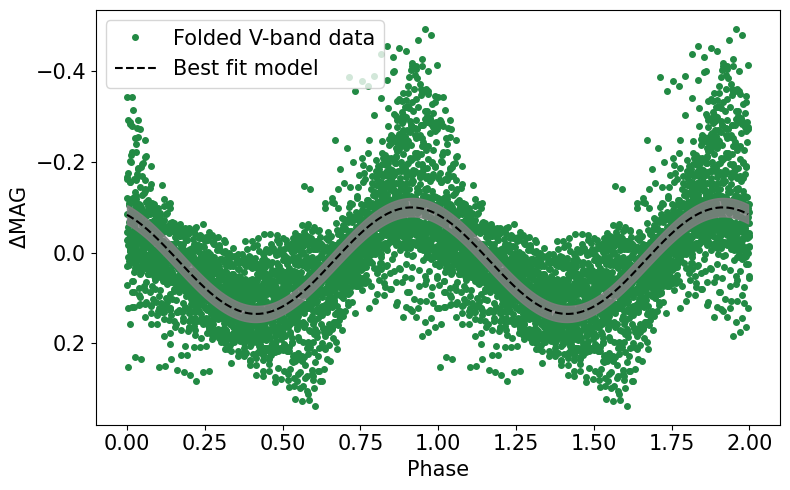}
    \caption{V-band data between MJD 58283 and MJD 58391 folded with the value of the frequency $f$ reported in Table \ref{tab:fit_results}. The black dotted line is the sinusoidal fitted model with the corresponding 3$\sigma$ confidence uncertainty shown as a gray shaded area. {Some residuals can be seen in the plot and are caused by the data within certain time intervals where the oscillation is not perfectly visible or slightly out of phase, as can be seen from Figure \ref{fig:phases_plots}.}}
    \label{fig:foldVband}
\end{figure}

\begin{figure*}
    \centering
    \includegraphics[width=1.\textwidth]{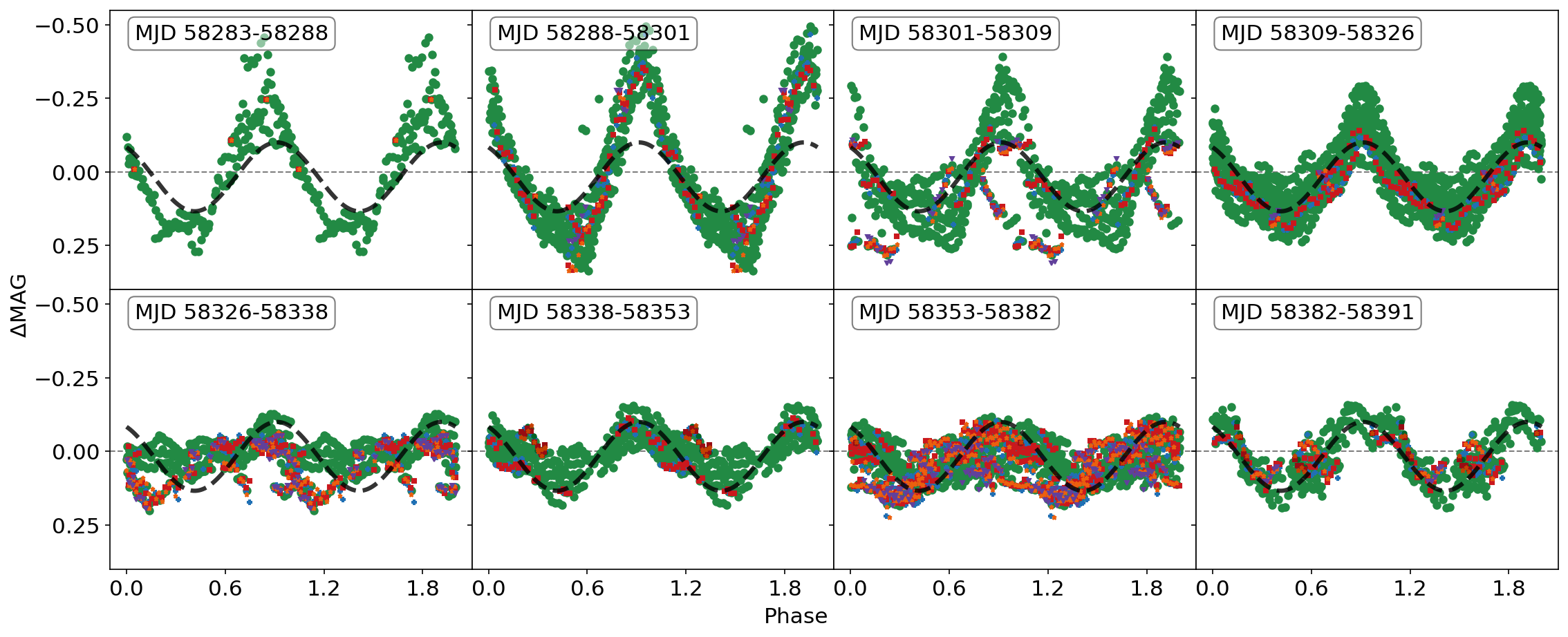}
    \caption{Evolution of the amplitude and shape of the periodical signal in the light curve of MAXI J1820+070 between MJD 58283 and MJD 58391 seen in the folded optical data in different sub-intervals. The colour code is the same of Figures \ref{fig:lc_longtime_not_corrected} and \ref{fig:cleaned_data}. The black dashed line corresponds to the best fitting sinusoid (equation \ref{eq:sinusoid}) for the entire interval. Except for the first panel, the phase of the signal remains constant, while its shape evolves with time. All the optical bands are in good agreement.}
    \label{fig:phases_plots}
\end{figure*}

Figure \ref{fig:foldVband} shows the V-band light curve folded using the frequency obtained with the MCMC fitting procedure together with the best fit model (black line) and the $2\sigma$ confidence interval (gray shaded area).  The overall behaviour is well described by the fitted profile, with only some outliers due to data within certain time intervals where the oscillation is not perfectly visible (or is slightly out of phase), as can be seen from Figure \ref{fig:phases_plots}.

We then detrended the data in the other bands using the same correction applied to the V-band data, properly rescaled for the different mean value of each band. The detrended data for the other bands are also shown in Figure \ref{fig:cleaned_data}. We tried to fit them using equation \ref{eq:sinusoid} but the measurements are too scattered to allow for the fitting procedure to converge to a single solution. Therefore, we tried to redo the fit by fixing the frequency to the value found in the V-band data and found that the phases in the different bands are in agreement with those in the V-band within the error bars.

The periodic signal shows a clear evolution of the amplitude and the shape. Looking at the light curves folded with the best-fitting frequency in different time sub-intervals, it is possible to see that the periodicity is not always significant, and in addition the shape of the profile and its amplitude change with time (Figure \ref{fig:phases_plots}). The first two sub-intervals (MJD 58283-58288 and MJD 58288-58301) show large amplitude oscillations -- even larger than the amplitude computed using the entire time interval -- and the first of these two sub-intervals seems to be slightly out of phase. In the following sub-intervals the periodicity is again clearly present, but with a lower amplitude and with some points out of phase (e.g. MJD 58301-58309 and MJD 58353-58382), while in other sub-intervals the periodicity seems to disappear completely (MJD 58326-58338). 

\begin{table}
    \centering
        \caption{Results of the fit of equation (\ref{eq:sinusoid}) to the optical and X-ray data.}
        \label{tab:fit_results}
    \begin{tabular}{lcc}
    \toprule\toprule
    Parameters & Optical & X-ray \\
    \midrule
        $t_0$ [MJD] & $58283.0$ & $58283.0$ \\ 
        $f$ [1/d]  & $1.4517\pm0.0001$ & $1.4517$ (fixed) \\
        $p$ [d]  & $0.68885\pm0.00005$ & $0.68885$ (fixed) \\
        $A$  & $0.117\pm0.002$ [MAG] & 240.7$\pm$15.5 [ct/s]\\
        $\phi$  [rad] & $0.67\pi\pm0.03$ & 1.77$\pi\pm$0.07\\
        $C$ & $0.017\pm0.002$ [MAG] & -17.5$\pm$11.4 [ct/s]\\
    \bottomrule
    \end{tabular}
    \tablefoot{$t_0$ is fixed, while the other parameters have been computed through a MCMC procedure \citep[\texttt{emcee;}][]{emcee2013}. The optical fit refers to the interval MJD 58283-58391, while the X-ray fit to the interval MJD 58309-58326. Since the X-ray data were not enough to compute an accurate value for the frequency $f$, we decided to fix it to the value found using the optical data and to free only the other parameters. For reference, we also report the period found by \protect\cite{Niijima2021} in a similar time window: $p_{_{N21}} = 0.688907\pm0.000009$ d.}
\end{table}

\subsubsection{X-ray light curve}

Similarly to what did for the photometric data, we tried to remove the overall short-term irregular oscillations and search for a hours-to-days periodic signal in the X-ray data. Before detrending the X-ray light curve, we rebinned the data in time to 60 s and subsequently renormalized them according to the number of active NICER focal plane modules, as the intense photon flux caused saturation of the internal telemetry in certain time intervals \citep{Homan2020}.
We applied the same detrending procedure described above, with the difference that we used 2 (instead of 3) days for the averaging of the data. This choice was dictated by the fact that the X-ray light curve shows more variability on a daily time scale. Moreover, we had to exclude from the detrending procedure the data in the MJD range 58302-58308 because of the intense X-ray flaring activity.

\begin{figure}
    \centering
    \includegraphics[width=1.\columnwidth]{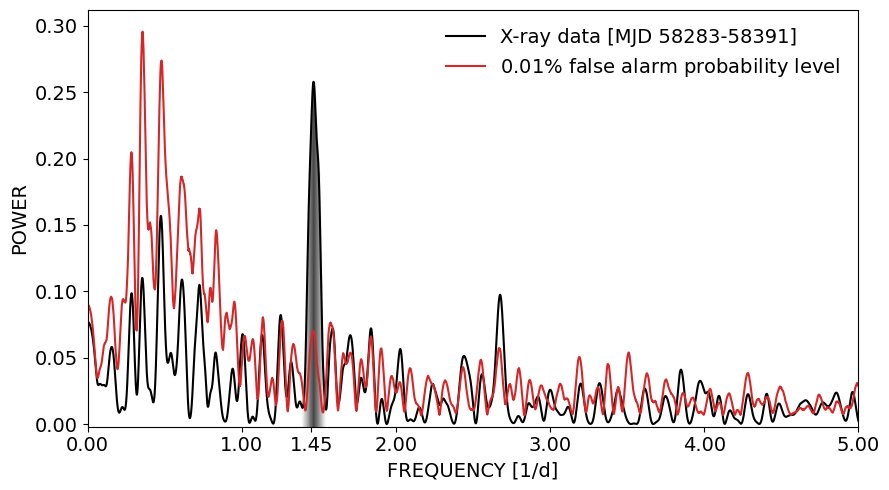}
    \caption{LS diagram (black curve) computed with the detrended X-ray data in the interval MJD 58309-58326. {The red curve correspond to the $0.01\%$ false alarm probability level (see text for details).} The highest peak is quite broad, but is centered at the frequency inferred from the optical photometry {and well above the false alarm probability level in the range of frequencies around it}.}
    \label{fig:xray_LS}
\end{figure}

\begin{figure}
    \centering
    \includegraphics[width=1.\columnwidth]{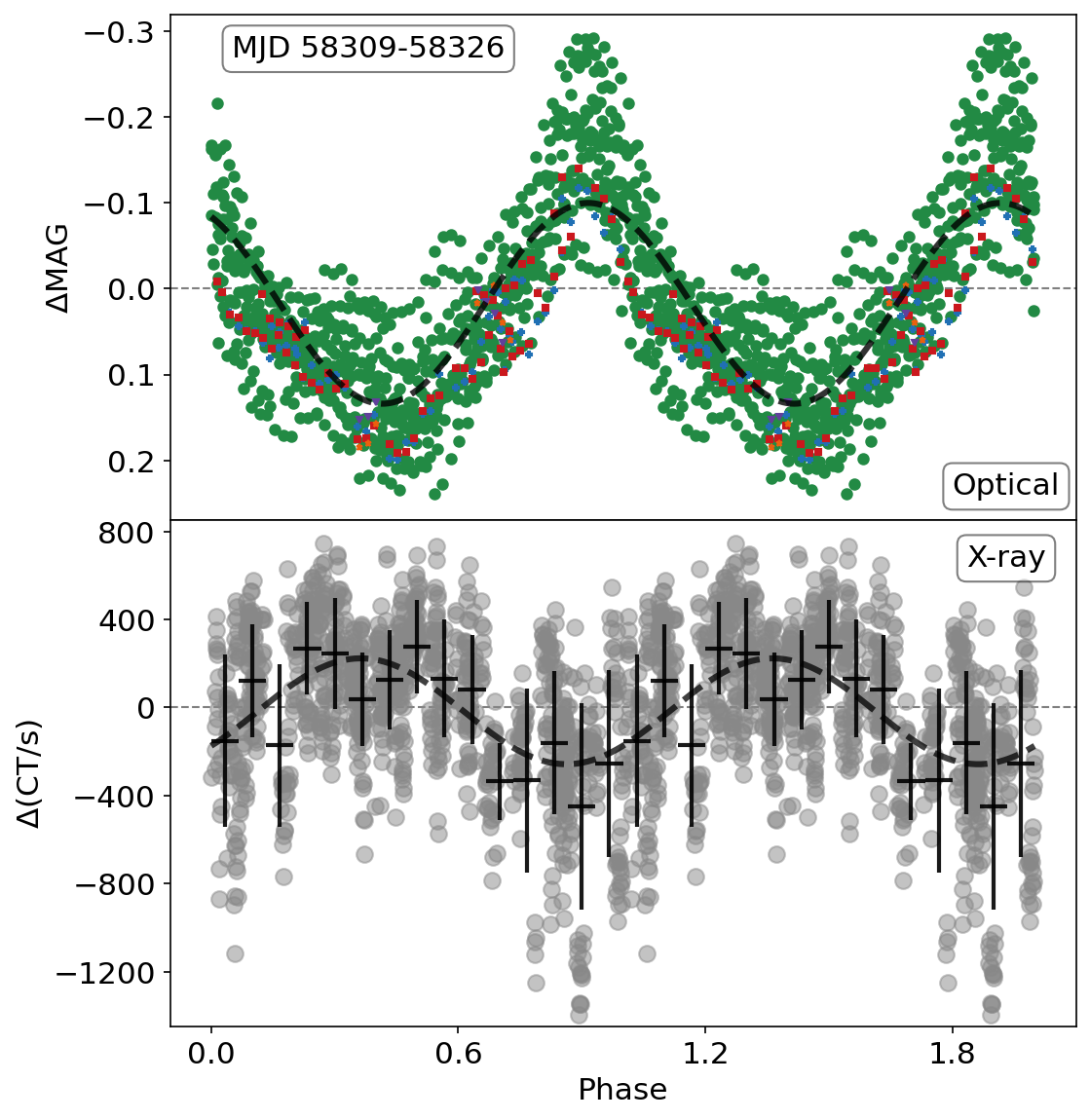}
    \caption{Comparison of optical and X-ray folded data in the interval MJD 58309-58326 computed using the best fitting frequency obtained with all optical data between MJD 58283 and MJD 58391 as reported in Table \ref{tab:fit_results}. Data are shown together with the best fitting sinusoid (black dashed lines). The data in the upper panel correspond to the optical data (coloured dots).  The lower panel shows the X-ray data (grey dots) as well as the same data binned in phase (black dots with error bars, 15 bins per phase). The error-bars are computed from the standard deviation of the data falling inside every phase bin.}
    \label{fig:xray_phased}
\end{figure}

After the detrending we found a small signal in a sub-interval of those in which we detected the optical periodicity (see Figure \ref{fig:xray_LS}). {We tested the possibility that the signal is caused by the data cleaning procedure by implementing an ad hoc bootstrap method (something similar to the method described in \citealt{lombscargle2017}): (1) we simulated many randomly distributed light curves by resampling the X-ray data but keeping the daily trend; (2) we applied exactly the same cleaning procedure described above; (3) we calculated all the LS diagrams and (4) we finally extracted the probability distribution at each frequency from which we could infer a false alarm probability level ($0.01\%$, red curve in Figure \ref{fig:xray_LS}).}  Remarkably, the resulting frequency is {out of the area where we could expect to measure spurious signals because of the cleaning process (frequencies $\leq1$ d$^{-1}$) and} close to the one found in the optical light curve in a time window where the optical periodicity is very well defined (MJD 58309-58326, upper right panel in Figure \ref{fig:phases_plots}). However, the significant variability of the X-ray light curve prevents us from effectively performing an accurate measurement. We then fixed the frequency at the value found in the optical data, leaving the phase, the amplitude, and the offset free to vary. We reported the results of the fit in Table \ref{tab:fit_results}. In Figure \ref{fig:xray_phased} we show the folded light curve together with the best-fit function. We also overplotted the same X-ray data binned in phase (the black points; 15 bins per phase) to better visualize the periodical trend in the X-ray data. {We measured a reduced $\chi^2_\nu$ value of 0.77 for the fitted sinusoidal model against a reduced $\chi^2_\nu$ value of 1.0 for a constant model, showing that a low significance periodicity is present in the data.}
{This periodicity, if real,} shows that the X-ray light curve is approximately in anti-phase with the optical light curve, with the X-ray leading the optical by about $1.1\pi$ radians or $\sim0.37887$ days ($\sim 9.1$ hours).

\subsection{Optical spectroscopy}

\begin{figure}
   \centering
   \includegraphics[width=1.\columnwidth]{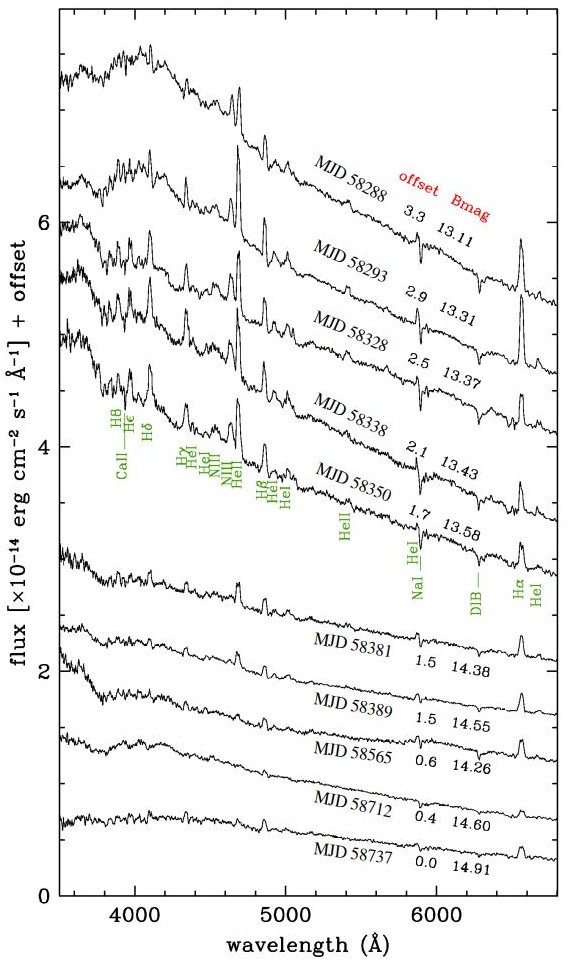}
      \caption{The 10 spectra of MAXI J1820+070 recorded with the Asiago 1.22m + B\&C
      telescope. From top to bottom, they are arranged in order of observing
      date. They have been offset for plot clarity by the indicated amount. 
      Next to the offset is listed the B-band magnitude exhibited by 
      MAXI J1820+070 at the time of the spectroscopic observation. The
      brightest emission lines are identified as well as a sample of the
      interstellar features seen in absorption (CaII, NaI, and the diffuse
      interstellar band at 6281~\AA\ as representative of several other
      well visible diffuse interstellar bands).}
         \label{fig:spectra_all}
   \end{figure}

The spectra presented in Fig. \ref{fig:spectra_all} are distributed over the first 5 optical maxima exhibited by MAXI J1820+070: the first 7 spectra cover the evolution from the 1st optical maximum to the 3rd optical maximum, the 8th spectrum is taken around the 4th optical maximum, and the last two spectra around the 5th optical maximum (the optical maxima are indicated with black dashed lines in Fig. \ref{fig:lc_longtime_not_corrected}). 
For a clearer view, the bluest part of the spectra (shortward of $3500$ \AA, which is much noisier than the rest) has been omitted from the Figure, and similarly for the wavelength interval red-ward of $6800$ \AA, where no significant emission lines are observed.

Irrespective of which maximum they belong, the slopes of the spectra in Fig. \ref{fig:spectra_all} can be clearly divided in two groups depending on the brightness of MAXI J1820+070: when the magnitude of the object is B$\leq$13.58, the spectral slopes are bluer ($<$$\ms{(B-R)}$$>=+0.33$) than those when the object turns fainter at B$\geq$14.26 ($<$$\ms{(B-R)}$$>=+0.42$). This change in the slope also corresponds to the time of the transition from the HS state to the IM state (and soon after to the LH state).
The continuum at $\lambda\leq3700$ \AA\ shows a great range of variability, which is unrelated to the flux in Balmer emission lines. This suggests that an interpretation in terms of a Balmer continuum going back and forth between emission and absorption seems unlikely.

\begin{figure}
   \centering
   \includegraphics[width=1.\columnwidth]{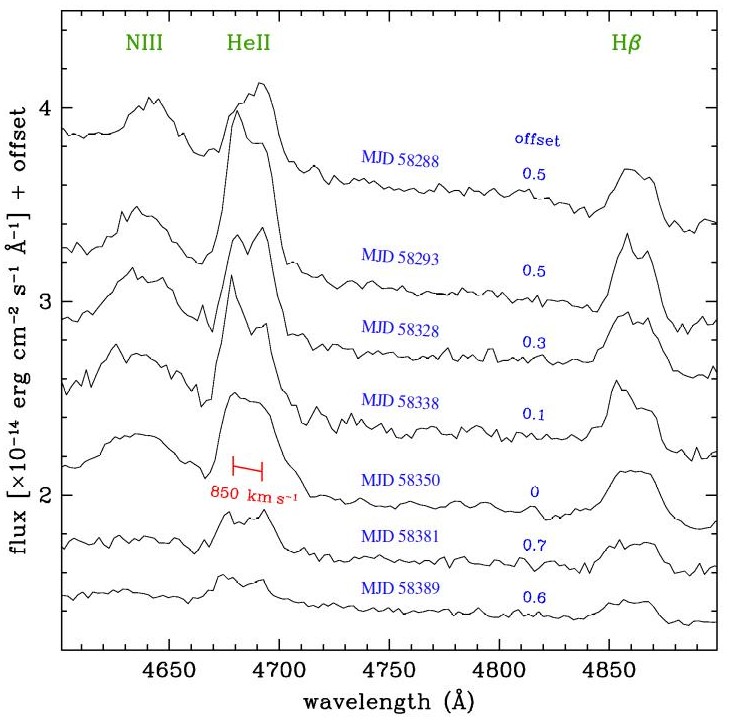}
      \caption{Zoomed view from Fig. \ref{fig:spectra_all} of the MAXI J1820+070 spectra for
      2018, covering hydrogen H$\beta$, HeII 4686~\AA\, and the broad
      4640~\AA\ blend attributed to NIII lines. The spectra are offset by
      the indicated quantity for clarity of the plot. The typical velocity
      separation of 850 km\,s$^{-1}$ for the double-peaked profiles is
      marked.}
         \label{fig:spectra_only2018}
   \end{figure}

Even if our spectra are characterized by a low-frequency resolution, the broad and double-peak profiles of emission lines are well resolved. The velocity separation between the blue and red peaks of the emission line profiles, and their relative intensity, vary according to the observing date and the specific line, with a mean value around $\sim$850 km\,s$^{-1}$ as illustrated in Fig. \ref{fig:spectra_only2018}.

\begin{figure}
  \centering
  \includegraphics[width=1.\columnwidth]{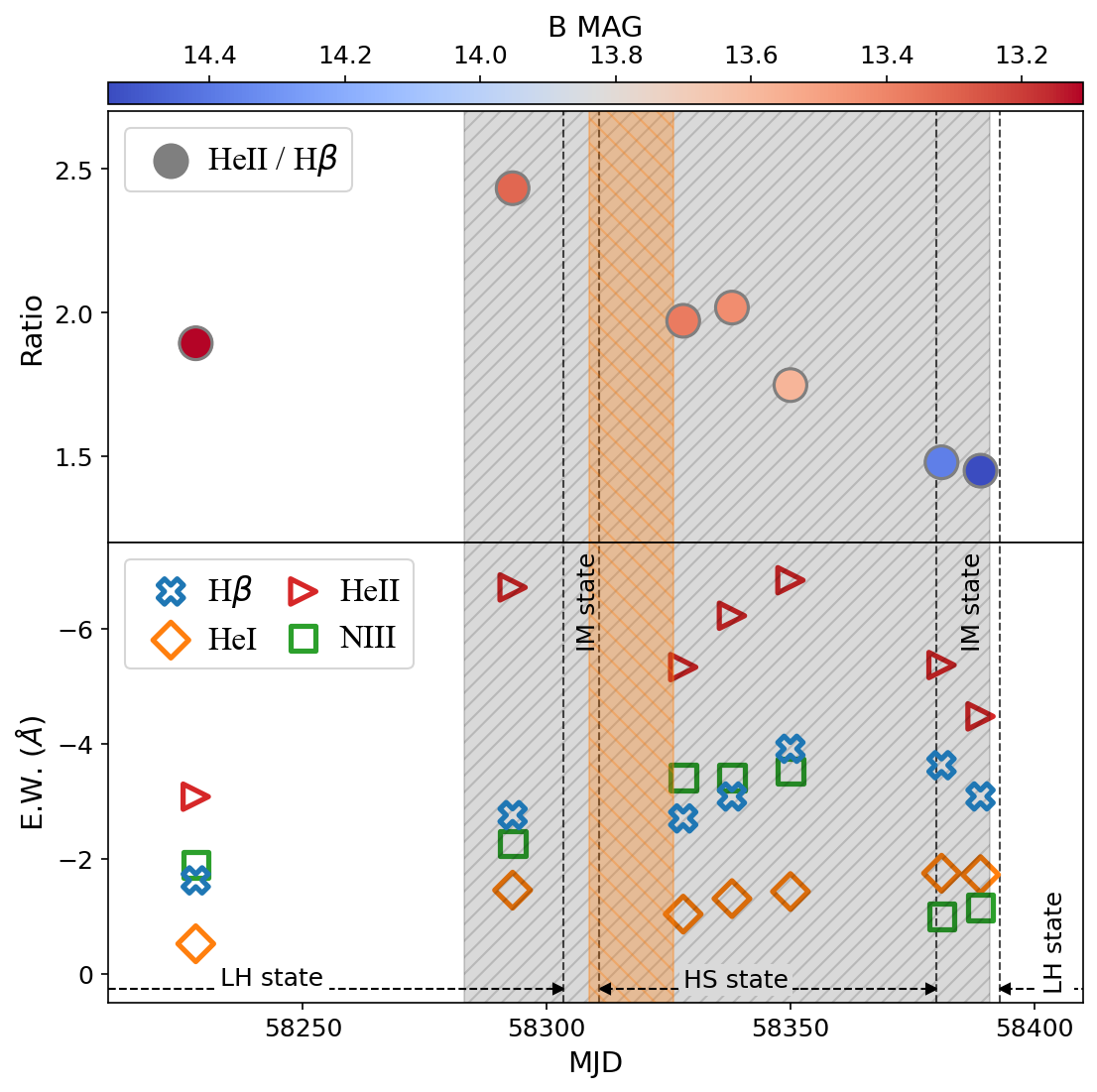}
      \caption{The lower panel shows the evolution of the equivalent width (in \AA) of representative emission lines as a function of time, calculated from the spectra of MAXI J1820+070 in Fig. \ref{fig:spectra_only2018}. The top panel shows the ratio between the equivalent widths of two of these lines, HeII 4686~\AA\ and H$\beta$.
      The colorbar on the top indicates the value of the B band magnitude of the source at the time of the observations. We show for comparison also the periods where the optical (gray shaded area) and the X-ray (orange shaded area) modulation are seen in the light curves, together with the epochs of the different accretion states that the source entered \citep{Shidatsu19}.}
         \label{fig:HeI}
  \end{figure}

The lower ionization/excitation emission lines, exemplified in Fig. \ref{fig:HeI} by H$\beta$ and HeI~5876~\AA, vary in phase with the level of continuum emission, which appear not to be the case for the higher ionization HeII 4686 \AA\ line. The equivalent width of the HeII line exceeds 2 times that of the H$\beta$ line for the earliest and brightest spectra in Fig. \ref{fig:spectra_only2018}, while it is 1.5 times of the H$\beta$ line or the later and fainter ones. A similar behavior is exhibited by the broad blend centered at 4640~\AA, evolving in parallel with HeII.  This suggests that the blend could be due to the \cite{Bowen1934} excitation mechanism, as observed in the high-density environments of novae and symbiotic binaries among other types of objects: HeII Lyman-$\alpha$ 303.78~\AA\ photons are absorbed by OIII in its ground state, that emits at 374.43~\AA\ upon returning to it.  The 374.43~\AA\ photons are absorbed by NIII in its ground state, and the following de-excitation produces a trio of lines around 4640~\AA\ (multiplet \#2 at 4634.16, 4640.64, and 4641.92~\AA).  In support to such a scenario, comes the abnormal large intensity of H$\delta$ in Fig. \ref{fig:spectra_all}, that violates the usual H$\beta$:H$\gamma$:H$\delta$:H$\epsilon$ progression.  The abnormal intensity would be easily explained by contribution from NIII lines (multiplet \#1 at 4097.31 and 4103.37~\AA) emitted during the same return to ground-state following the pumping by absorption of 374.43~\AA\ photons emitted by OIII. 
The 4640~\AA\ blend is also prominent in the spectra obtained by \cite{MunozDarias2019} at epochs preceding our first spectroscopic observation.

\subsection{Analysis of the high timing resolution data}

Measuring the variability of a signal on different time scales can be achieved efficiently by moving to the Fourier space (as for the LS diagrams in the previous section) and calculating the Power Density Spectrum (PDS) to measure the variance of a signal at each Fourier frequency and pinpoint the presence of periodic or semi-periodic features in the light curves.
We then searched for fast periodic features in the NICER X-ray data and the IQ/AQ+ optical data by calculating the PDS using Fast Fourier Transform (FFT) algorithms. 

We computed the average PDS for each observing run, both in the X-ray and optical band, with a frequency resolution that varies from dataset to dataset (typically between 5 mHz and 20 mHz) and with a constant Nyquist frequency of 0.5 kHz (all light curves are time-binned at 1 ms). The frequency resolution is not constant because it depends on the width of the time window into which the observations are segmented before performing the FFTs, typically of the order of $\sim50$, $\sim100$ or even $\sim200$ seconds.  All segments are finally averaged to compute the final PDS for each observing run. For very long exposures it was possible to use $\sim200$ seconds, as an adequate number of time frames were available to obtain an average PDS of good quality. In the case of shorter exposures, however, we had to use time frames of 50/100 seconds. The PDS are normalized according to \cite{Leahy1983} so that the Poisson noise level is about 2. We verified that at higher frequencies (>100 Hz) the PDS are all flat and close to 2. 

Once the PDS were calculated we used \textsc{xspec} \citep{XSPEC1996} to model the data. We followed the method described in \cite{Ingram2012} to transform the PDS into the correct format needed for \textsc{xspec}, where we could easily fit the data with some functional models. In all PDS, we fitted the data using multiple components: a power law for the Poissonian noise at high frequencies, a zero-centred Lorentzian profile for band-limited noise, and a series of wide and narrow Lorentzian profiles for the different features in the PDS. 
In this way, we were able to evaluate the properties of the QPOs in the X-ray and optical bands.\\

\begin{figure}
    \centering
    \includegraphics[width=1.\columnwidth]{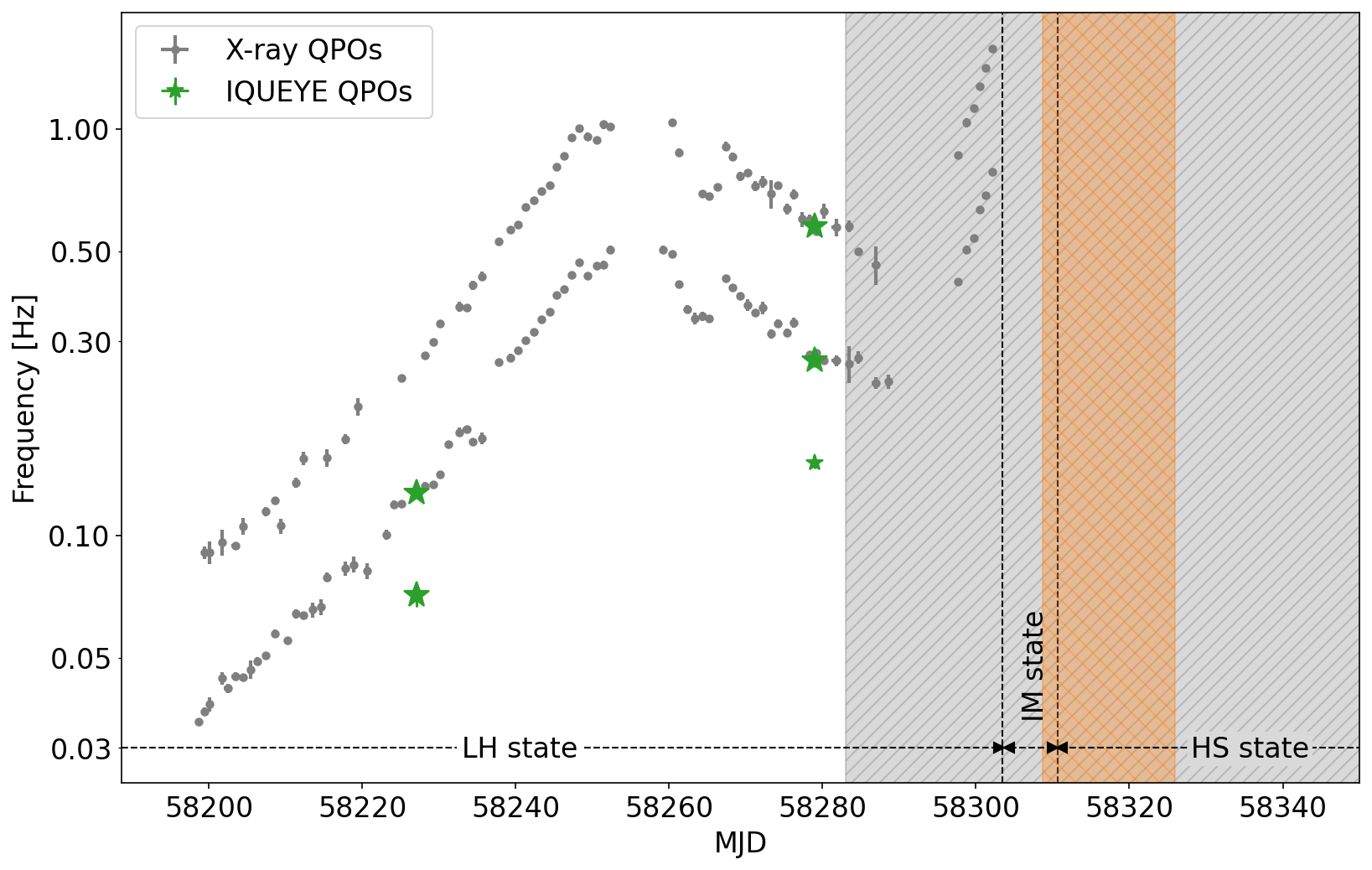}
    \caption{NICER QPOs central frequencies evolution (gray points) compared to the optical QPOs (green stars) and the periods where the optical (gray shaded area) and the X-ray (orange shaded area) modulation can be seen in the light curves. We also show the epochs of the different accretion states that the source entered, as reported by \citet{Shidatsu19}.}
    \label{fig:QPOs_overallEVO}
\end{figure}

\begin{figure*}
    \centering
    \includegraphics[width=1.\columnwidth]{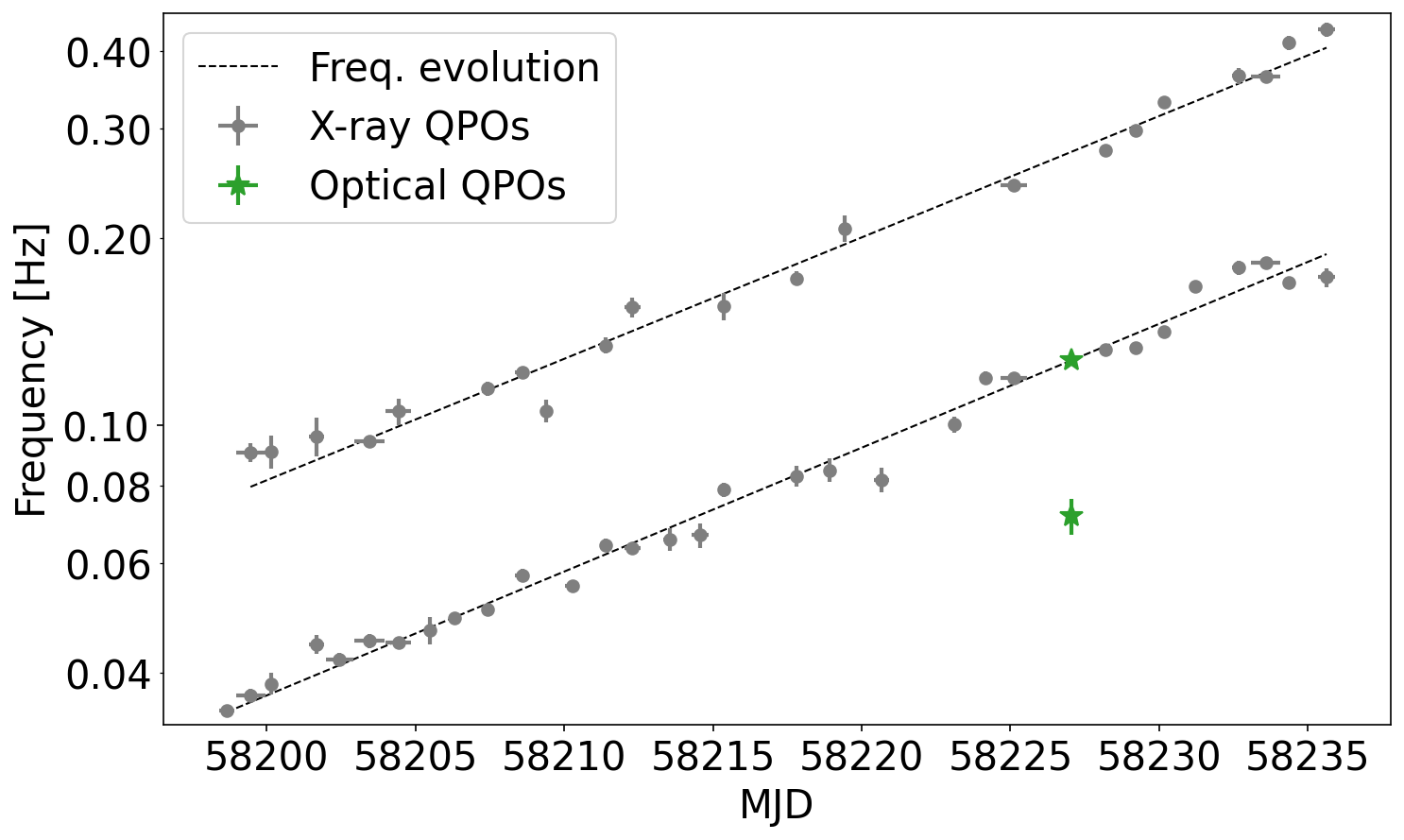}
    \includegraphics[width=1.\columnwidth]{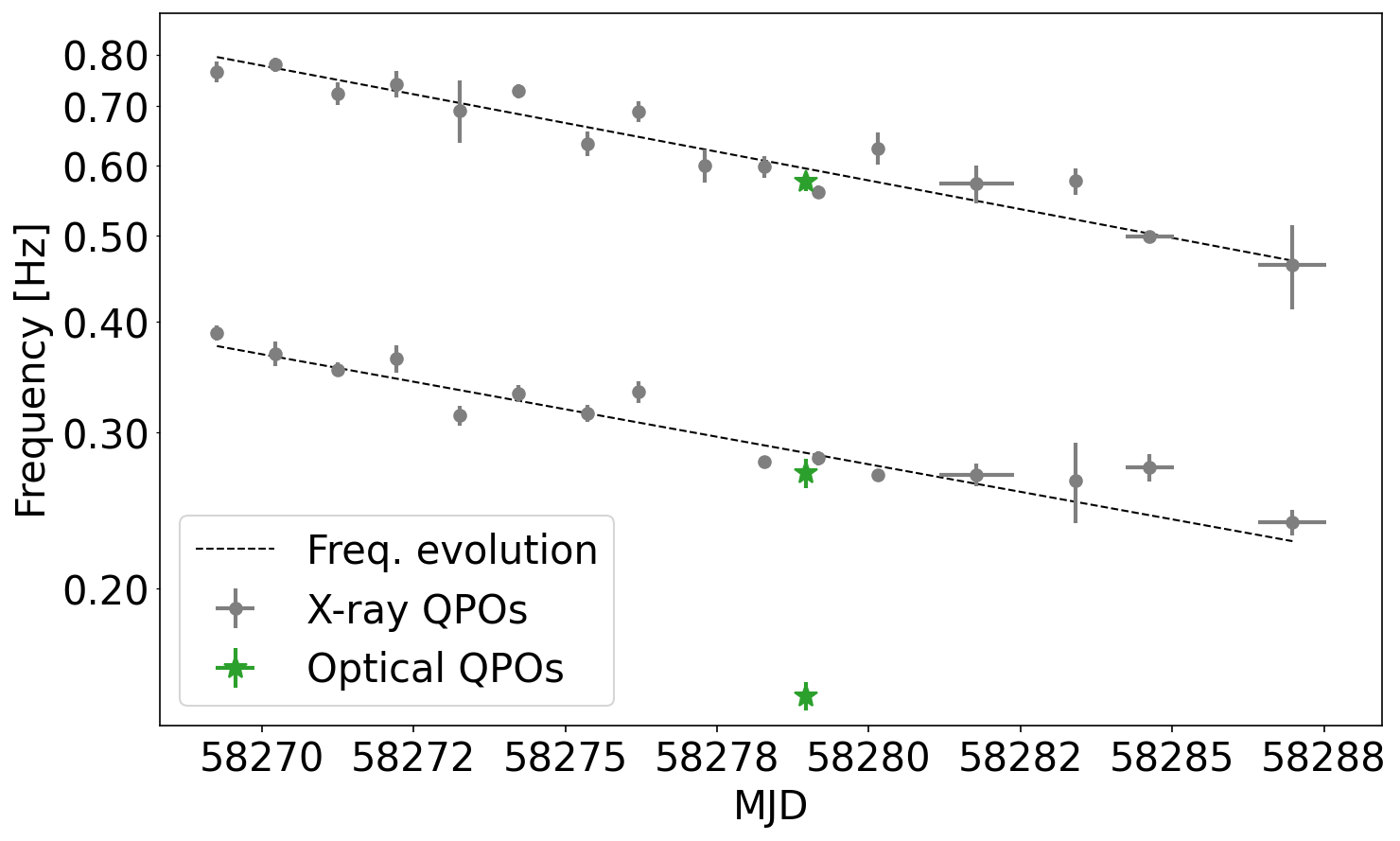}
    \caption{Time evolution of the central frequency of the two most prominent X-ray QPOs in NICER data (grey points) together with the central frequency of the optical QPOs (green stars). On the left we show the evolution during March-April 2018, while on the right that during June 2018.}
    \label{fig:qposEVO}
\end{figure*}

\begin{figure*}
    \centering
    \includegraphics[width=1.\columnwidth]{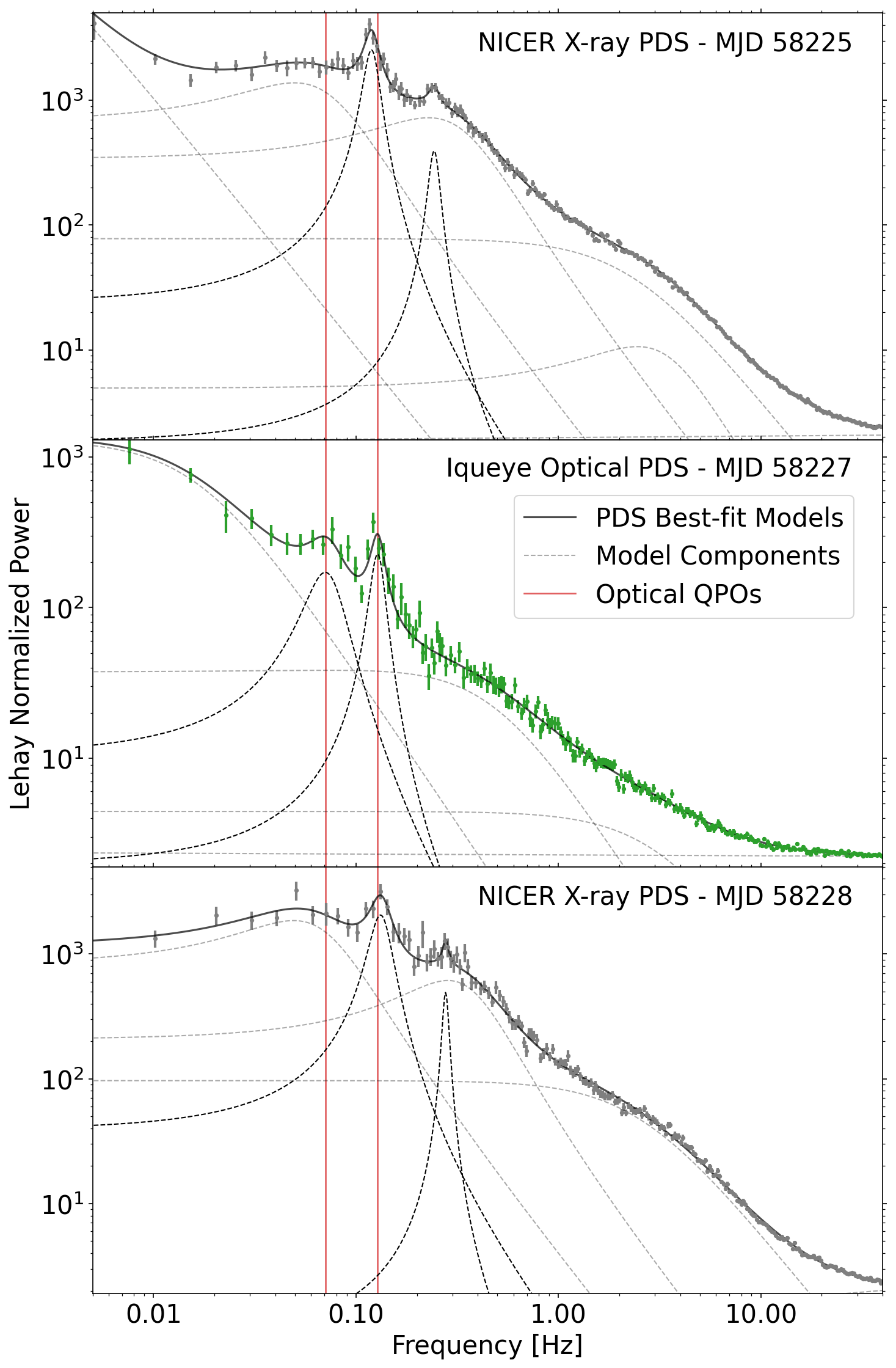}
    \includegraphics[width=1.\columnwidth]{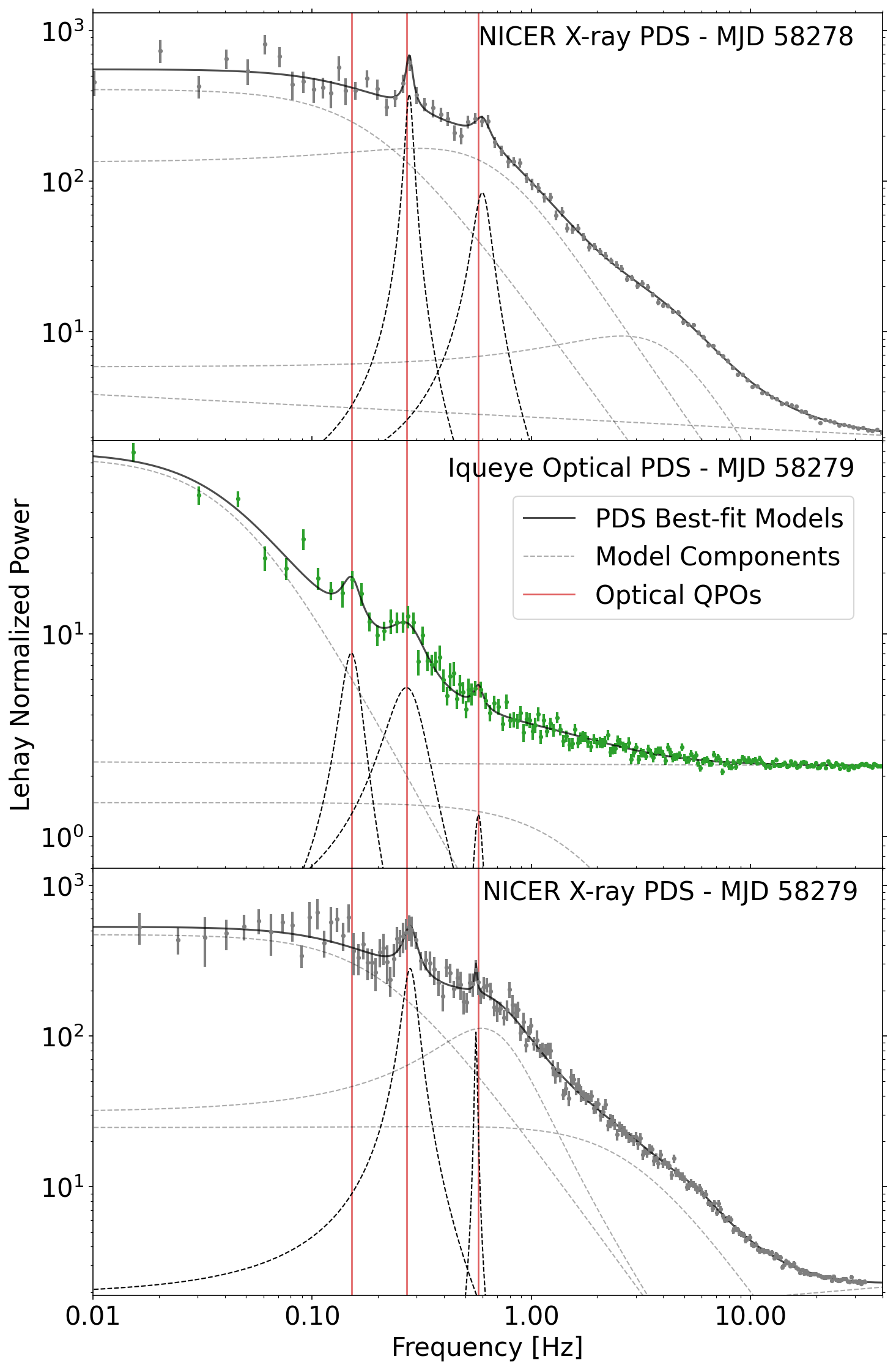}
    \caption{Comparison of optical and X-ray PDS of MAXIJ1820+070. The top and bottom panels show the NICER PDS while the middle panels show the Iqueye PDS.  The left panels correspond to the data taken around 18 April 2018 and those on the right to the data taken around 8 June 2018. As no overlapping observations are available, we used for comparison the X-ray observations taken before and after the optical observations. The vertical red lines correspond to the central frequency of the QPOs {observed} in the optical PDS. In both periods the frequency of the most prominent QPO in the X-rays is in good agreement with the frequency of the second QPO in the optical band.}
    \label{fig:qposOptvsX}
\end{figure*}

\begin{figure}
    \centering
    \includegraphics[width=1.\columnwidth]{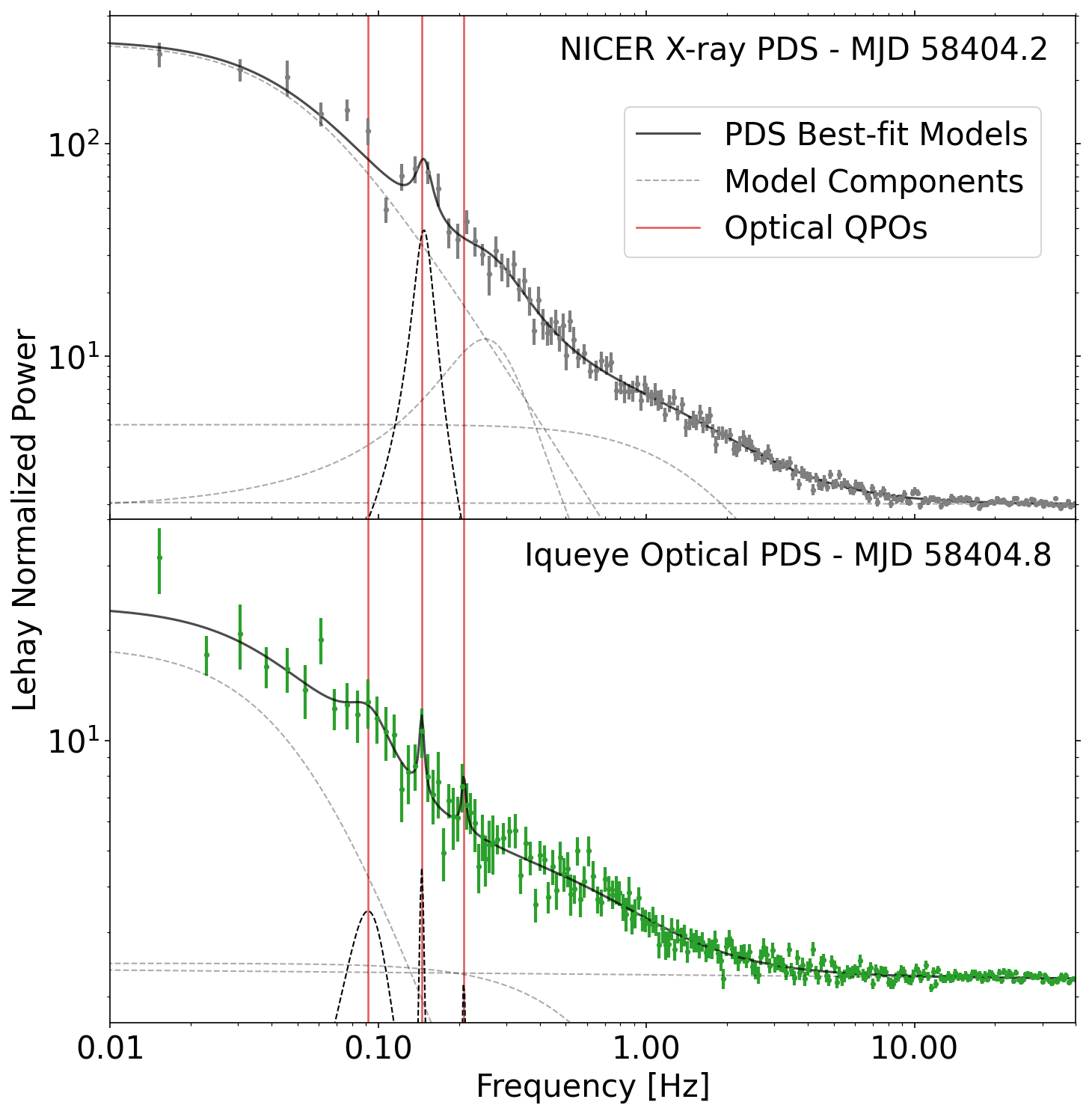}
    \caption{Comparison of optical and X-ray PDS of MAXIJ1820+070 for the data taken around 13 October 2018. The top panel show the NICER PDS and the bottom panel show the Iqueye PDS.  As no overlapping observations are available, we used for comparison the closest X-ray observation to our optical observation. The X-ray observation after our optical observation shows no clear QPO and therefore we do not show the corresponding PDS. The vertical red lines correspond to the central frequency of the QPOs in the optical PDS. The frequency of the most prominent optical QPO (the second one in the optical PDS) is in good agreement with the frequency of the X-ray QPO.}
    \label{fig:qposOptvsX2}
\end{figure}

\begin{table*}\centering
\begin{threeparttable}
\caption{{Central frequency ($\nu_{_0}$), full-width-half-maximum (FWHM) and significance ($\sigma$) of the optical LFQPOs observed in three different epochs (April, June and October 2018) and shown in Figures \ref{fig:qposOptvsX} and \ref{fig:qposOptvsX2}.}}\label{tab:QPO_freq}
\begin{tabular}{lrrcrrcrrc}\toprule\toprule
&\multicolumn{3}{c}{{April 2018}} &\multicolumn{3}{c}{{June 2018}} &\multicolumn{3}{c}{{October 2018}} \\
&\multicolumn{3}{c}{{[MJD 58227]}} &\multicolumn{3}{c}{{[MJD 58279]}} &\multicolumn{3}{c}{{[MJD 58404.8]}} \\\midrule
&\multicolumn{1}{c}{$\nu_{_0}$} & \multicolumn{1}{c}{FWHM} & {$\sigma$\tnote{*}} & \multicolumn{1}{c}{$\nu_{_0}$} & \multicolumn{1}{c}{FWHM} & {$\sigma$\tnote{*}} & \multicolumn{1}{c}{$\nu_{_0}$} & \multicolumn{1}{c}{FWHM} & {$\sigma$\tnote{*}} \\
\midrule
{QPO1} & $ 71\pm 4$ mHz & $ 36\pm16$ mHz & 2.7 & $151\pm 6$ mHz & $ 37\pm16$ mHz & 2.6 & $ 92\pm 9$ mHz & $ 45\pm43$ mHz & 1.0 \\
{QPO2} & $128\pm 2$ mHz & $ 24\pm 5$ mHz & 5.2 & $269\pm10$ mHz & $132\pm32$ mHz & 4.6 & $145\pm 4$ mHz & $  7\pm 6$ mHz & 1.5 \\
{QPO3} &                &                &     & $575\pm13$ mHz & $ 68\pm51$ mHz & 1.8 & $209\pm 4$ mHz & $  9\pm14$ mHz & 1.3 \\
\bottomrule
\end{tabular}
\begin{tablenotes}
       \item [*] The significance of the LFQPOs is given by the ratio between the normalization of the Lorentzian fitted to the QPO and its uncertainty (as in \citealt{Motta2015}).
       \end{tablenotes}
\end{threeparttable}
\end{table*}

From the PDS of the optical observations, we found QPOs in three observations: in April, June and October 2018 (Figures \ref{fig:qposOptvsX} and \ref{fig:qposOptvsX2} and Table \ref{tab:QPO_freq}). 
To exclude spurious signals due to systematic effects in the optical data, we tried to search for similar components in the PDS of the reference star and we did not find any hint of similar features. It is worth noting that the total fractional rms of the reference star during the April and June sessions was always substantially lower than the total fractional rms of MAXI J1820+070. (see Tab. \ref{tab:log_hightimeres}).

In the X-rays, the property of the QPOs have been already extensively analyzed by \cite{Stiele20}. 
A series of type-C LFQPOs are detected in the \textit{Swift}/XRT and NICER data soon after the beginning of the first outburst (MJD 58198).
These type-C QPOs have frequencies between 30 mHz and $\sim1-2$ Hz, and are seen until the source enters the IM state (from that moment on type-B QPOs appear in the PDS). 
Type-C QPOs reappeared again in the NICER data in October 18 after the new transition to the LH state.
We did a similar analysis of the NICER data to compare the X-ray PDS to the optical one. We then focus our analyses in the periods close to the Iqueye/Aqueye+ observations (for example we did not study in detail the phase near the state transitions and during all the HS state, roughly between the beginning of July and September 18). In Fig. \ref{fig:QPOs_overallEVO} we show the overall evolution of the central frequencies ($\nu_{_0}$) of the two most prominent QPOs found in the NICER data until the first state transition\footnote{Sometimes the lower frequency component that we see in the optical data is marginally {visible} also in the X-ray data, but always with a much smaller normalization. We thus decided not to include it in Fig. \ref{fig:QPOs_overallEVO}}. We also show the epochs of the different accretion states that the source entered and the epochs where it was possible to see the modulation of the light curve in the optical data (gray shaded area) and in the X-ray data (orange dashed area). An evolution of the central frequencies of the QPOs is clearly visible.\\

We divided the data into 7 intervals, corresponding to different evolving characteristics of the QPOs central frequencies.

\textit{March-April 2018}: In the optical, in MJD 58227, we observed two prominent QPOs \citep{QPO_Apr18} on the top of three broad-band noise components (see middle-left panel of Figure \ref{fig:qposOptvsX} and Table \ref{tab:QPO_freq}). Instead, in almost all the NICER data in the MJD range 58198-58236, we can see two QPOs (a main component and an upper harmonic; see Fig. \ref{fig:qposEVO}). 
The central frequencies of these two components evolve exponentially with time with the same trend ($\nu_0 \propto 10^{0.02\cdot t}$). The main QPO in the optical data (upper green star) perfectly fits the expected frequency found from the X-ray trend. In the left panels of Fig. \ref{fig:qposOptvsX} we show a comparison of the X-ray/optical PDS considering the X-ray observations taken before and after the optical observation. We see a close match between the central frequencies of the main components, as well as the presence of a lower frequency component that is visible only in the optical data. The upper harmonic is clearly visible in the X-ray data.

\textit{April-May 2018}: The evolution of the central frequencies in this interval (MJD 58236-58249) is similar to the previous one, but with a slightly different dependence with the time ($\nu_0 \propto 10^{0.025\cdot t}$). No optical observations are present in this interval.
    
\textit{May 2018}: During May 2018 (MJD 58249-58268), the frequencies of the QPOs stop increasing. In MJD  58259, we did observe the source with Iqueye, but we did not see any QPOs. This is most probably caused by the bad weather, since in the NICER observations before and after our observation the two QPOs are clearly visible. We estimated that a QPO with properties similar to the QPOs found in April or June had to have a fractional rms variability smaller than $\sim2\%$ (95\% confidence level, \citealt{QPO_Apr18}) for the QPO to not be observable.
    
\textit{June 2018}: In the optical, in MJD 58279, we observed again two prominent QPOs \citep{QPO_Jun18} on the top of two broad-band noise component, plus a small third component that can be identified only when comparing the PDS of the optical and X-ray observations (see middle-right panel of Figure \ref{fig:qposOptvsX} and Table \ref{tab:QPO_freq}). From the X-rays, in the interval MJD 58268-58288 we see that the frequencies decrease exponentially with time (Fig. \ref{fig:qposEVO}, right panel) and again the trends for the main component and the upper-harmonic are the same ($\nu_0 \propto 10^{-0.013\cdot t}$). The main QPO in the Iqueye data and the upper-harmonic fit perfectly the expected frequencies found in the X-rays. Looking again at Fig. \ref{fig:qposOptvsX} we see a close correspondence between the central frequency of the main QPO in the X-ray and optical bands and a lower frequency component, which is again only visible in the optical data. The upper-harmonic, clearly visible in the X-ray data, is only marginally {observed} in the optical data.

\textit{June-July 2018}: Before the state transition to the IM state (at $\sim$ MJD 58303.5) the evolution of the central frequencies of the QPOs changes again. At first the QPOs almost disappear, with only the upper-harmonic marginally visible in some of the observations (as reported by \citealt{Stiele20}). We decided not to include these very low-amplitude QPOs in Fig. \ref{fig:QPOs_overallEVO}. Around MJD 58297, the two harmonically related QPOs become well {visible} again with an exponentially increasing central frequency. Finally, when the source enters the IM state, a quickly evolving type-B QPO is observed in the NICER data (not shown in Fig. \ref{fig:QPOs_overallEVO}; more details for this phase can be found in \citealt{Homan2020}). No optical observations are present in this interval.
    
\textit{July to September 2018}: After the state transition and during all the period when the source was in the HS state no QPOs are found. This is again in agreement with the optical data, where no QPOs are present. During July, when we also observed with Aqueye+ and the meteorological conditions were good, we estimated a fractional variability smaller than $\sim0.5\%$ (95\% confidence level, \citealt{QPO_Jul18}) for a QPO not to be observable.
     
\textit{October-November 2018}: When the source entered again the LH state, QPOs started to reappear in the NICER data and in the optical data. In the optical, in MJD 58404, we observed other three QPOs on top of broad-band noise components (see bottom panel of Figure \ref{fig:qposOptvsX2} and Table \ref{tab:QPO_freq}). In the X-rays, the QPOs can be seen only in a few observations (as again reported by \citealt{Stiele20}). Moreover, there is no clear evolving trend as observed at earlier times.

\section{Discussion}\label{sec:discussion}

\subsection{Optical super-orbital modulation}

\begin{figure}
    \centering
    \includegraphics[width=1.\columnwidth]{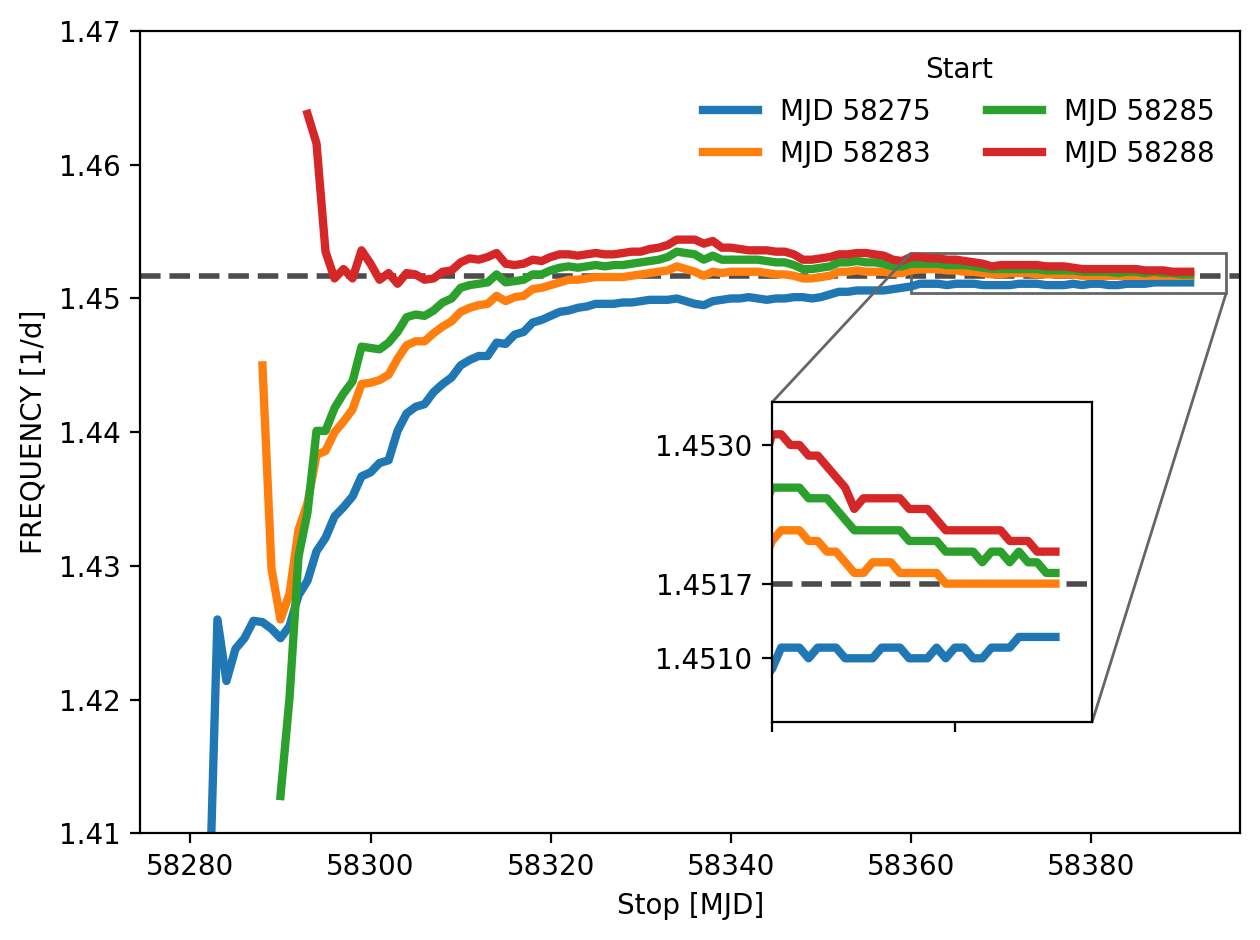}
    \includegraphics[width=1.\columnwidth]{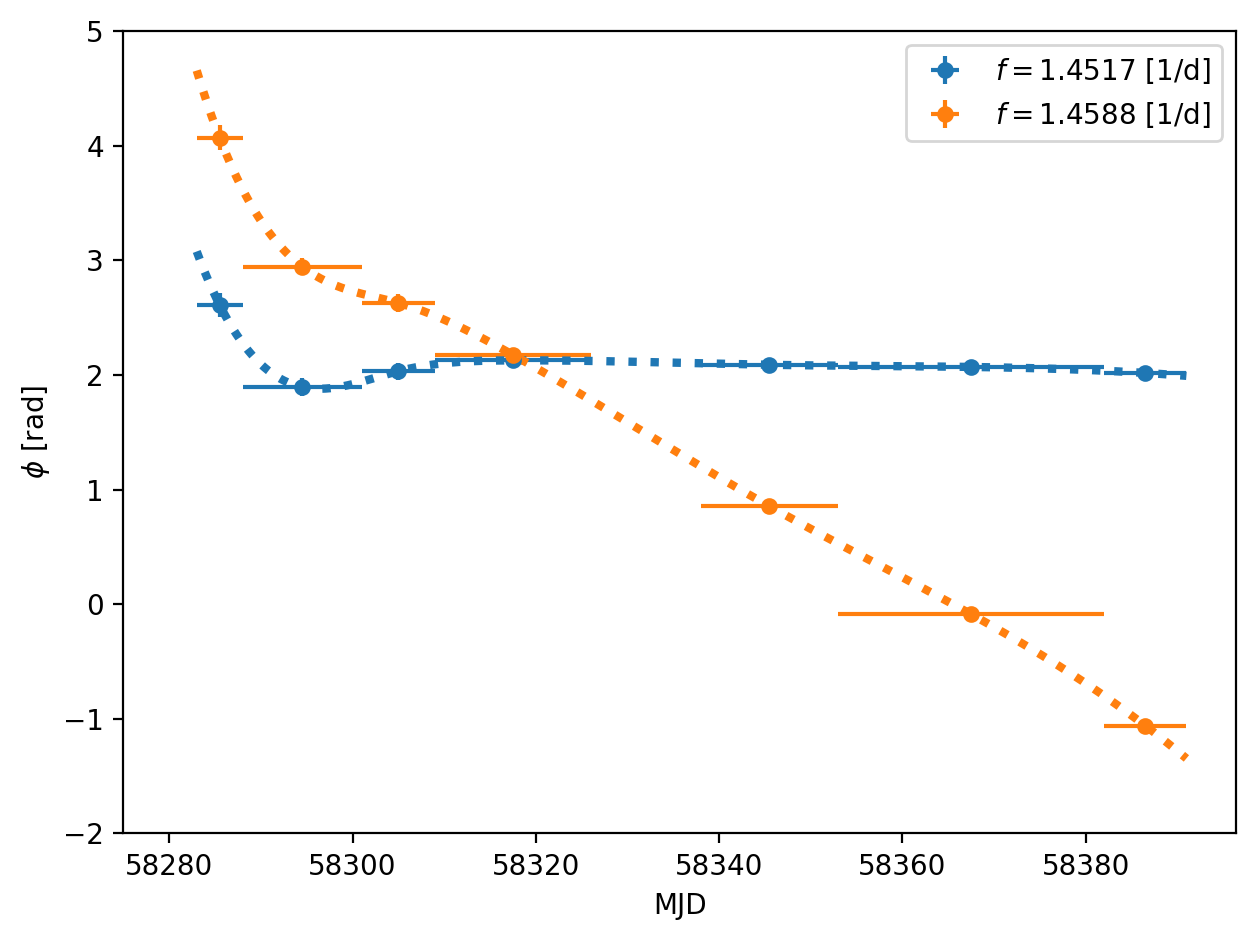}
    \caption{\textit{Top:} Frequency of the highest peak in the LS diagrams computed considering different starting dates. The $x$-axis correspond to the stopping date used for computing the LS diagrams. The frequency always quickly evolves to the reference value of $\sim1.4517$ $1/d^{-1}$ (black dashed line). In addition, as the starting date is moved forward, the curves tend faster and faster towards the reference value. The curve that deviates most from this trend is the one corresponding to the starting date MJD 58275 (blue curve), which includes data when the modulation was still evolving. \textit{Bottom}: Phase evolution, considering the dates shown in Fig. \ref{fig:phases_plots} (excluding the data in MJD 58326-58338), using two different frequencies. In blue the frequency computed in this work ($f_{_{58283-58391}} = 1.4517$ 1/d) and in orange the frequency related to the orbital motion ($f_{_{orb}} = 1.4588$ 1/d, \citealt{Torres2019}). The dotted lines are cubic splines fitted to the data.}
    \label{fig:diff_mjd_start}
\end{figure}

{The values of the periodicities mentioned in this Section are reported in Table \ref{tab:DifferentPeriods}.}
From the optical photometry, we calculated a period ($p_{_{58283-58391}}$) that is about $0.5\%$ longer then the orbital period ($p_{_{orb}}$) found from optical spectroscopy in quiescence \citep{Torres2019} and about $2\%$ shorter than the one reported by \citealt{patterson18} ($p_{_{P18}}$). Later \citet{Patterson19} reported a slightly shorter period ($p_{_{P19}}$) considering a different time window. 
The latter two values were calculated over an interval of about 15 and 30 days, which only partially overlaps with the interval used in our analysis. We repeated the analysis in similar intervals and we measured two periods that are within $1\sigma$ to those in \cite{patterson18} and \cite{Patterson19} ($p_{_{58275-58290}}$ and $p_{_{58275-58310}}$).
We also found that adding or removing dates at the beginning or at the end of this interval, greatly changes the inferred modulation period and leads to a higher actual uncertainty on this measurement. To understand these differences we have studied the behaviour of the inferred period using different starting and ending dates. This is shown in the upper panel of Fig. \ref{fig:diff_mjd_start}. The $y$-axis shows the values of the frequencies of the peak in the corresponding LS diagrams. We can see that, when we add more and more data, the frequency always evolves to a value close to the reference value of $f_{_{58283-58391}} = 1.4517$ 1/d.  In addition, as the starting date is moved forward (going from MJD 58725 for the blue curve to MJD 58288 for the red curve), the different curves tend faster and faster towards the reference value. The blue curve, i.e. the one including data at earlier dates (MJD $<58283$), seems to deviate more from the reference value, both considering the inferred frequency using only the early data (MJD $<58309$) and the whole time window (up to MJD 59391, zoomed-in panel in Fig. \ref{fig:diff_mjd_start}). Looking at the blue curve, it is clear that the period computed using MJD 58275-58290 and MJD 58275-58310 is longer than the one inferred from all the data between MJD 58283-58391. This may be caused by the fast evolution of this signal in the first few days after its appearance.

\begin{table*}\centering
\begin{threeparttable}
\caption{{Comparison of the super-orbital periods that can be inferred using different temporal windows.}}\label{tab:DifferentPeriods}
\begin{tabular}{lllll}\toprule\toprule
{Name} & {Period [d]} & {Freq. [1/d]} & {Window [MJD]} & {Source} \\\midrule
$p_{_{orb}}$ [$f_{_{orb}}$]\tnote{**} & 0.68549(1) &1.45881(1) & &\cite{Torres2019} \\
$p_{_{58283-58391}}$ [$f_{_{58283-58391}}$] & 0.68885(5) &1.4517(1) &58283-58391 &This work \\
$p_{_{N21}}$ [$f_{_{N21}}$] & 0.688907(9) &1.45157(2) &58283-58391\tnote{*} &\cite{Niijima2021} \\
$p_{_{P18}}$ [$f_{_{P18}}$] & 0.703(3) &1.422(6) &58275-58290\tnote{*} &\cite{patterson18} \\
$p_{_{58275-58290}}$ [$f_{_{58275-58290}}$] & 0.701(1) &1.426(2) &58275-58290 &This work \\
$p_{_{P19}}$ [$f_{_{P19}}$] & 0.6903(3) &1.4486(6) &58275-58310\tnote{*} &\cite{Patterson19} \\
$p_{_{58275-58310}}$ [$f_{_{58275-58310}}$] & 0.6920(5) &1.4451(9) &58275-58310 &This work \\
\bottomrule
\end{tabular}
\begin{tablenotes}
       \item [*] These dates are only estimated from plots and thus can be different from the effective dates used to compute the periods in the cited papers.
       \item [**] True orbital period reported for comparison.
\end{tablenotes}
\end{threeparttable}
\end{table*}

To check the validity of our assumption, we tried to fit the frequency and the phase of the modulation in different time intervals, following what was done by \cite{ThomasSupehump2021}. We first fitted the frequency and the phase dividing the data in two intervals, before and after MJD 58309. Considering the data in MJD 58283-58309, we found a frequency of $1.4488\pm0.0005$ 1/d and a phase of $0.72\pi\pm0.02$ ({consistent within $1\sigma$ with} the value reported in \citealt{Patterson19}), while considering the data in MJD 58309-58391 we found a frequency of $1.4514\pm0.0003$  and a phase of $0.72\pi\pm0.03$ ({consistent within $1\sigma$} with the frequency found in the entire time window). 
The phase of the signal, varying the frequency in the two time windows, remains constant. 

We then fit the phase of the signal fixing the frequency at specific values: the frequency $f_{_{58283-58391}}$ and the orbital frequency $f_{orb}$ reported by \cite{Torres2019}. 
The phases are fitted considering the same time windows of Fig. \ref{fig:phases_plots} and are plotted in the bottom panel of Fig. \ref{fig:diff_mjd_start}. 
Using $f_{_{58283-58391}}$ (blue points) we found a roughly constant phase in all the windows -- only the data in the first window are slightly out of phase -- and a good agreement with the overall phase ($\phi_{_{58283-58391}} = 0.67\pi\pm0.03$). 
On the other hand, using $f_{_{orb}}$, we found that the phase monotonically decreases with time, starting from $\sim1.3\pi$ down to $\sim-0.3\pi$. 
This can be easily interpreted as being caused by an incorrect value of the frequency, so that in order to compensate for it, the phase must evolve over time.

From all these findings we therefore tentatively conclude that the modulation is not caused by the orbital motion (as suggested by \citealt{ThomasSupehump2021} for all the data after MJD 58309) and that it probably evolved from the frequencies reported in \cite{patterson18} and \cite{Patterson19} to the frequency $f_{_{58283-58391}}$. A similar result is also reported in \citeauthor{Niijima2021} (\citeyear{Niijima2021}; see the top panel of their Fig. 2), who found a period {within $2\sigma$} to the one reported here ($p_{_{N21}}$; Table \ref{tab:DifferentPeriods}) and interpreted it as the true period of the {super-orbital modulation}.

\subsection{X-ray/optical super-orbital modulation}

A clue that the optical modulation visible during MJD 58283-58391 and the X-ray modulation visible during MJD 58309-58326 are driven by the same physical mechanism is given by the behaviour exhibited during the different state transitions.
The optical modulation started to show up around the beginning of the first optical rebrightening, as can be seen in Fig. \ref{fig:lc_longtime_not_corrected}.
At that time MAXI J1820+070 was still in the LH state. However, around the same dates, the evolution of the QPOs central frequency in the X-ray band changed (Fig. \ref{fig:QPOs_overallEVO}). The optical modulation then lasted all along the HS state and stopped only few days after the new transition to the LH state. However, the modulation of the X-ray light curve appeared only immediately after the state transition from the IM to the HS states and lasted only for a few days. Interestingly, the optical modulation immediately after MJD 58326 (when the X-ray modulation is no longer visible) also seems to vanish for a few days (bottom left panel of Fig. \ref{fig:phases_plots}) before appearing again, but with a much smaller amplitude. 
Indeed, this could be the proof that the process that generates the optical modulation could be the same that produces the modulation that seems visible in the X-rays between MJD 58309 and MJD 58326 (Fig. \ref{fig:xray_phased}).

As already reported, {if the X-ray modulation between MJD 58309 and 58326 is real},the optical and X-ray are out of phase by about $\sim1.1\pi$, with the X-ray leading the optical. Given the large uncertainty in the X-ray modulation and that in a small time window, we can also fit the optical data using the orbital modulation and compare the phases in the two bands using the orbital solution by \cite{Torres2019} (using the orbital frequency for both bands or only for the X-ray band). Even with these assumptions the X-ray always leads the optical, again by about $\sim1.1\pi$. These evidences appear to be consistent with a common geometrical origin of the optical and X-ray modulations (e.g. a warped disc).

\subsection{X-ray/optical LFQPOs}

\begin{figure*}
    \centering
    \includegraphics[width=1.\columnwidth]{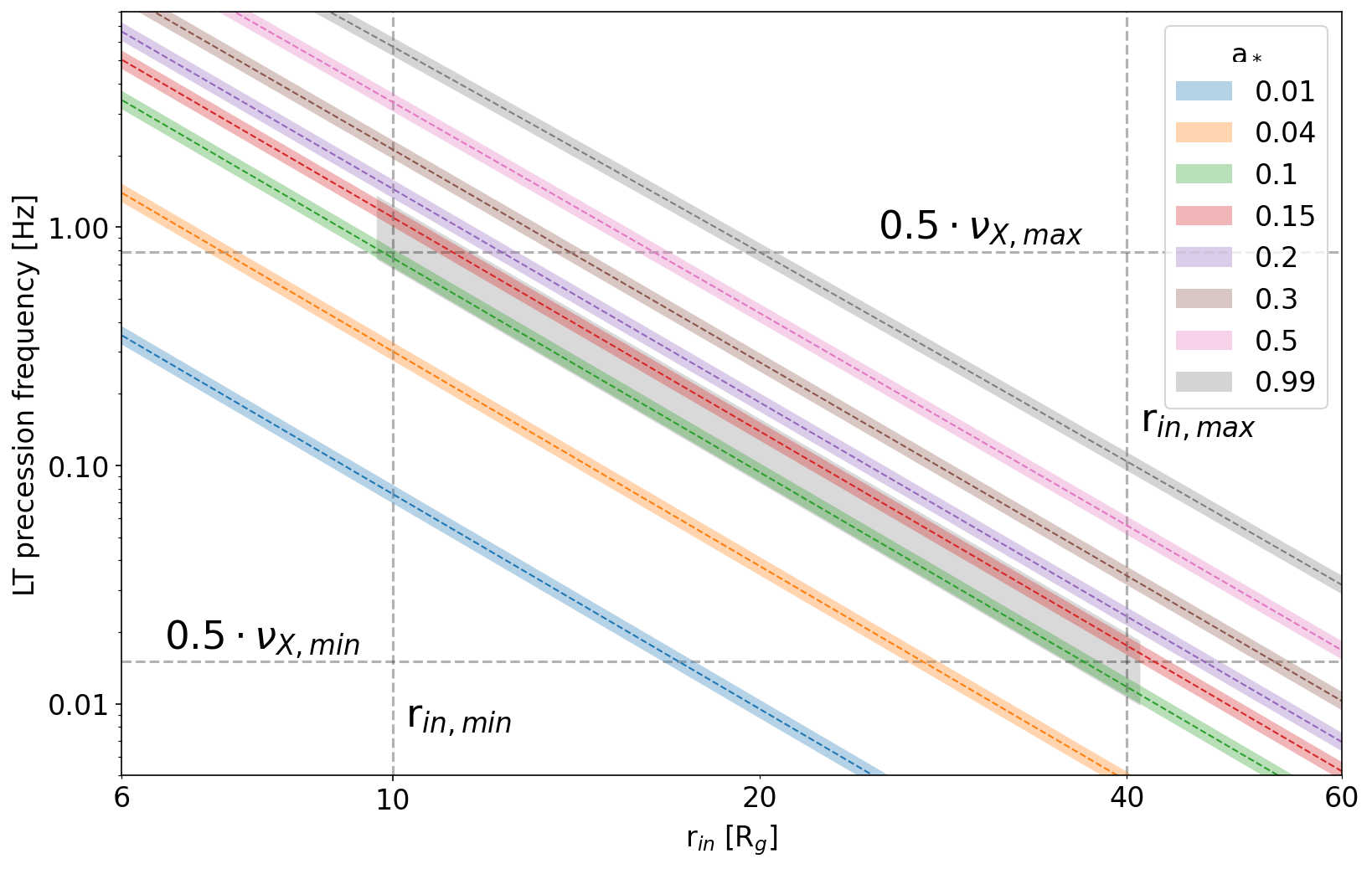}
    \includegraphics[width=1.\columnwidth]{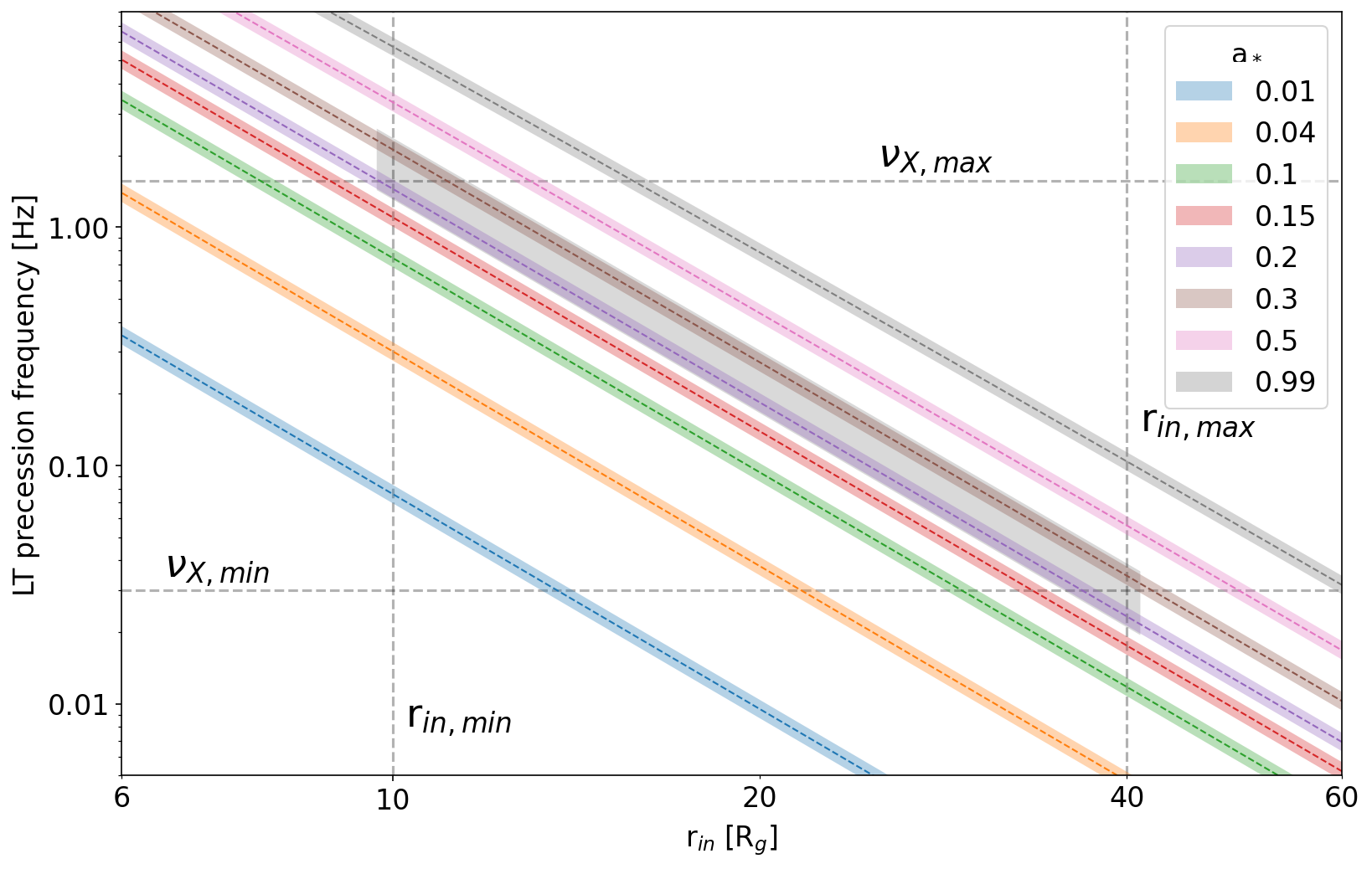}
    \caption{LT precession frequency computed following eq. (1) in \citealt{Ingram2009}, as a function of the truncation radius r$\rs{in}$ for different values of the black-hole spin a$_*$ (coloured lines).
    The vertical lines correspond to the estimated range of radii for the truncation radius, while the horizontal lines correspond to the minimum and maximum LT precession frequency that we considered. 
    The shaded coloured areas around the lines are the uncertainties from estimated mass of the black-hole as reported by \citealt{Torres2020}, while the gray area represents the parameter space that is in agreement with the data. On the left we are showing the solutions that correspond to the scenario in which the fundamental frequency is traced by the QPOs at lower frequencies seen in the optical data.
    On the right instead we are showing the solutions that correspond to the scenario in which the fundamental frequency is traced by the most prominent peak in the X-rays.}
    \label{fig:LT}
\end{figure*}

The type-C LFQPOs {observed} in the PDS are synchronous over at least 5 orders of magnitude in terms of energy, from the optical band up to the hard X-ray band \citep{Zampieri2019, MA2021, Mao2022, ThomasQPOs2022}.
A close link between the optical and X-ray bands is also evident from the PDS obtained in this work (see Figs. \ref{fig:qposOptvsX} and \ref{fig:qposOptvsX2}), even if no strictly simultaneous X-ray/optical observations were carried out. The frequencies of the LFQPOs in the two bands are consistent once the expected evolution of the central frequency is taken into account (Figs. \ref{fig:qposOptvsX}, \ref{fig:qposOptvsX2}, \ref{fig:QPOs_overallEVO}, \ref{fig:qposEVO}). This result is confirmed by other authors who analyzed simultaneous optical/X-ray data \citep{Paice2019, Paice2021, ThomasQPOs2022}. Remarkably, the optical QPOs are present in all our observations when the source was in the LH state. In two observations out of three it was possible to detect a lower frequency component in the optical PDS (in October 2018 the lower frequency component is probably present, but seems to be very broad), while in the X-rays this component is not observed. The frequencies of the lower component in the optical data and of the most prominent peak in the X-rays appear to be harmonically related, with a ratio of $\sim$1:2. On the other hand, a higher frequency component is almost always {present} in the X-ray PDS, while in optical data it is {always less pronounced}.

Optical, ultraviolet and infrared QPOs have been observed in other black hole binaries in the past \citep{Motch1983, Imamura1990, Hynes2003, Durant2009, Gandhi2010, Veledina2015, Kalamkar2016, Vincentelli2019, Vincentelli2021}. In several cases the central frequency of the QPOs is the same in the different bands \citep{Hynes2003, Veledina2015, Vincentelli2021}. In other cases there was no clear identification of an X-ray counterpart, either because no simultaneous X-ray observations were available \citep{Imamura1990} or because it was just not detectable \citep{Durant2009, Gandhi2010, Vincentelli2019}.
Interestingly, in GX 339-4, the central frequency of the optical/infrared QPO is half of the frequency of the QPO in the X-ray band \citep{Motch1983, Kalamkar2016}.
Usually the fundamental QPO is considered to be the one with the highest peak amplitude in the X-ray band \citep{IngramMotta2019} {based on the fact that in many periodic processes the fundamental frequency has higher amplitude than its overtones. However,} \cite{Veledina2013} predicted that if the inclination of the system is between 45° and 75°, the second harmonic could be stronger than the fundamental.
Therefore, depending on the adopted model, the first harmonic of the QPO could be the one {observed} in the optical/infrared/ultraviolet energy range.

Regardless of whether or not the fundamental frequency is that observed in the optical band, how can we explain the close link between X-ray and optical emission? We have already seen that there are evidences that the regions where the photons in the two bands are emitted must be very close \citep{Paice2019, Paice2021}.  {However, the geometry of the system is yet to be fully understood and therefore different models can be adopted to interpret the data. Assuming a lamp-post geometry for the hot corona in the innermost regions of the accretion flow, measurements of the reverberation time lags between the X-ray continuum emission and the (broad) iron K emission line (produced in the outer irradiated accretion disc) imply the existence of an accretion disc with a characteristic size $6-20$ times smaller (15 gravitational radii, $\ms{R}\rs{g}$) than what previously seen in other similar objects \citep{Kara2019}. Given that the reverberation time lag frequency increases with time, the hot corona should also contract and decrease its height following the evolution of the outburst \citep{Kara2019, Buisson2019}.
A different interpretation however is possible, considering a standard disc model \citep{ShakuraSunyaev1973} where the steady increase of the reverberation lag frequency is caused by the decrease of the internal truncation radius $\ms{r}\rs{in}$ \citep{DeMarco2021, Kawamura2022}. This interpretation is also supported by broadband X-rays spectra that show that $\ms{r}\rs{in}$ should be located close to the central black hole, at only a few tens of gravitational radii \citep{Marino2021}.}

We can thus try to model the properties of the observed LFQPOs by considering a Lense-Thirring (LT) precession of the hot accretion flow near the black hole \citep{Stella1998, Motta2014}. The precession could occur either in the hot inner flow \citep{Ingram2009, Veledina2013} or at the base of the jet \citep{Markoff2005, Stevens2016}. 
{In fact, the observed LFQPOs may originate from the precession of a small-scale jet, as suggested by the gradual increase of the soft phase lag with increasing energy, up to and above 200 keV \citep{MA2021}.} 
The evolution of the frequency of the LFQPOs (Fig. \ref{fig:QPOs_overallEVO}) implies a variable characteristic radius and, since the frequency is small ($<1$ Hz), this radius can be relatively large (larger than the innermost stable circular orbit). In the LT precession model, the precession frequency ($\nu\rs{LT}$) is generally linked to the frequency of the fundamental QPO. {Assuming that the fundamental frequency corresponds to the centroid frequency of the QPO in the optical} ($\nu\rs{opt}$), at half of the frequency of the most prominent peak in the X-ray data ($\nu\rs{X}$), implies that the precession frequency is half of what is inferred from the X-rays. The weak/lack of detection of the first harmonic in the X-rays may be caused by, e.g., partial obscuration/eclipse of the X-ray emitting region at certain precession phases. This is in agreement with the prediction of \cite{Veledina2013} given that the inclination of the system is estimated to be $\sim63$° \citep{Atri2020}. Thus, considering that the range of the most prominent QPO in the X-ray band is between $\sim$30 mHz and $\sim$1570 mHz \citep[maximum frequency reached before the transition to type-B QPO;][]{Stiele20}, the precession frequency should then be between $\nu\rs{LT,min}\simeq15$ mHz and $\nu\rs{LT,max}\simeq785$ mHz.

We can try to estimate the dimensionless spin of the central black-hole, a$_*$, using the LT precession model. From the expression of the Lense-Thirring precession frequency of a particle in the Kerr metric \citep[e.g.][]{Merloni1999, Ingram2009}:
\begin{equation}\label{eq:LTprecess}
    \nu\rs{LT} = \nu\rs{\phi}\biggl[1 - \sqrt{1 - \frac{4\ms{a}_*}{\ms{r}^{3/2}}+\frac{3\ms{a}^2_*}{\ms{r}^2}}\biggr],
\end{equation}
where $\nu\rs{\phi}$ is the frequency of a circular orbit in the equatorial plane at a given radius $\ms{r}$ (in units of the gravitational radius $\ms{R}\rs{g}=GM/c^2$).
Here, the radius is the truncation radius of the standard disc $\ms{r}\rs{in}$.
In the following, we assume that it varies in the range $10-40$ $\ms{R}\rs{g}$, accordingly to the findings of \cite{DeMarco2021} and \cite{Kawamura2022}. {If} we assume that the minimum and maximum precession frequencies are linked to the maximum and minimum radii at which the standard disc is truncated at a certain time, from a comparison of eq. \eqref{eq:LTprecess} with the observed frequencies, we derive an estimate for a$_*$.\footnote{{Clearly, all the reported estimates of a$_*$ are significantly dependent on the chosen interval of $\ms{r}\rs{in}$}.}
The results are shown in Fig. \ref{fig:LT}, where we considered two different scenarios: the scenario in which the true fundamental frequency is the one {visible} in the optical data (left panel) and the scenario in which the true fundamental frequency is the most prominent peak seen in the X-rays (right panel). The two panels show the LT precession frequency as a function of the truncation radius for different values of the black-hole spin (coloured lines). The vertical lines correspond to the estimated range of radii for the truncation radius, while the horizontal lines correspond to the minimum and maximum LT precession frequency that we considered. The gray area represents the parameter space that is in agreement with the data.
Taking into account the uncertainties on $\ms{r}\rs{in}$, if the fundamental frequency is the one observed in the optical band, the allowed dimensionless spin of the black-hole should be $\lesssim 0.15$, meaning a relatively slowly spinning black hole. This value is in good agreement with the spin estimated by \cite{Zhang2020}.
Conversely, if the fundamental frequency is the one observed in the X-ray band, the allowed a$_*$ should be $\lesssim 0.3$, meaning a black hole that can spin twice faster then in the previous scenario.
It is worth emphasising that any estimate of the black hole spin depends strongly on both the adopted model and the choice of the inner disc truncation radius. Clearly, different assumptions may lead to different results  \citep[see e.g.][]{Bhargava2021}. In particular, although the LT precession model provides a reasonable physical framework for producing the observed QPOs frequencies, different interpretations of the reverberation frequency lag and/or the geometry of the corona are possible \citep{Kara2019, Buisson2019, DeMarco2021, Kawamura2022}, making the estimate of the truncation radius somewhat uncertain. However, as mentioned above, selecting $\ms{r}\rs{in}$ in the range inferred from some recent X-ray spectral analyses \citep{DeMarco2021, Kawamura2022} we obtain consistent results.

We note that a small value of the inner disc radius ($\sim$10 $\ms{R}\rs{g}$) is also consistent with the onset of the super-orbital modulation at the end of the LH state, because the stronger X-ray irradiation from the inner disc can more easily trigger the warp observed in the X-rays and in optical.

\subsection{Optical spectroscopy}

From the optical spectroscopic data we obtain additional interesting information on the system during the first phase of the outburst. The blue spectra up until MJD $\sim$58380 are indicative of an increased X-ray irradiation of the outer disc. Indeed, until MAXI J1820+070 returns to the LH state, the X-ray flux is high (see e.g. Fig. 1 in \citealt{Prabhakar2022}) and irradiation is significant. Furthermore, when the X-ray flux is maximum, the ambient density is probably high enough to lead to the activation of the Bowen mechanism, as observed in the broad blend centred at 4640 \AA. 
Finally, the variability of the profile of some broad lines (e.g. HeII 4686 \AA) may be a further indicator of a (possibly warped) precessing disc. When the  super-orbital modulation starts to be detected, the equivalent width of the spectral lines in Fig. \ref{fig:HeI} tends to increase, while it decreases when the modulation vanishes. Therefore, these spectral features seem to disappear when the super-orbital modulation stops and the first phase of the outburst terminates. This is a further indication that there exists a close relationship between the emission in the X-ray and optical bands.

\section{Conclusions}\label{sec:conclusions}

In this work we presented a detailed analysis of data obtained from an intensive multi-timescale variability observing campaign of the new black hole X-ray binary MAXI J1820+070.
We made use of low-cadence and fast optical photometry data, as well as NICER X-ray data. The main part of our analysis is focused on the initial phases of the outburst, from March to October 2018. 

We detected an optical super-orbital modulation that started around the beginning of the first optical rebrightening with a frequency $f=1.4517\pm0.0001$ 1/d and an evolving profile.
After the transition from the LH to the HS state, a hint of a modulation seems to emerge for a few days also in the X-ray band. We found evidence of a X-ray super-orbital modulation with a frequency in close agreement with the optical one and out of phase by about $\sim1.1\pi$ (X-ray leading the optical).
The optical super-orbital modulation lasted until the transition from the HS to the LH state.

In three epochs (April, June and October 2018) we {observed} synchronous optical and X-ray type-C LFQPOs.
A lower frequency component, harmonically related to the X-ray most prominent QPO (with a ratio 1:2), is {visible} in two optical observations.
If the lowest modulation frequency is that observed in the optical power density spectrum, the characteristic variability frequency of MAXI J1820+070 is lower than that inferred from the `fundamental' QPO in the X-rays. 
Considering the Lense-Thirring precession model, under the assumption that the truncation radius of the standard disc varies in a range $\ms{r}\rs{in} \simeq 10-40$ $\ms{R}\rs{g}$, and taking as precession frequency the lowest frequency traced by the optical observations, we estimated a relatively slowly spinning black-hole with a$_*\lesssim0.15$, in close agreement with other measurement in the literature.
In these assumptions the observed optical and X-ray large scale super-orbital warp could be triggered by the self-irradiation of the outer disc induced by the inner disc when extending inward ($\ms{r}\rs{in} \sim 10$ R$\rs{g}$), before the source enters the HS state.

Further observations of black-hole binaries in outburst at different wavelengths and on multiple time scales can help explain how synchronous QPOs are produced over such a large energy range (from the optical to the hard X-rays) and understanding if the super-orbital modulation is directly linked to the process that generates the QPOs.  
\begin{acknowledgements}
    This research made use of the following PYTHON packages: MATPLOTLIB \citep{Matplotlib2007}, NUMPY \citep{Numpy2011}, SCIPY \citep{Scipy2020}, PANDAS \citep{Pandas2010}, ASTROPY \citep{astropy:2013, astropy2018, astropy:2022} and EMCEE \citep{emcee2013}. We acknowledge financial support from INAF Research Grant "Uncovering the optical beat of the fastest magnetized neutron stars (FANS)". (S.S.) The study was conducted under the state assignment of Lomonosov Moscow State University and also supported by grants from the Slovak Academy VEGA 2/0030/21 and APVV-20-0148.
\end{acknowledgements}

\bibliographystyle{aa}
\bibliography{MAXIJ1820_bib} 

\onecolumn
\begin{appendix}
\section{Log of the observations}
In this appendix we are reporting the summaries of the optical observations. We do not report the summary of the NICER X-ray observations since it can be retrieved from the official HEASARC archive\footnote{\url{heasarc.gsfc.nasa.gov/docs/nicer/nicer_archive.html}}.

\begin{table*}[!th]
\centering
\tiny
\caption{Summary of the optical fast timing observations.}
\label{tab:log_hightimeres}
\begin{tabular}{llllrrrr}
    \toprule\toprule
 Night &   Telescope ID & Start (MJD) & Stop (MJD) &  T$_{obs}$ (s) &  Count rate (ct/s)  &  Frac. RMS (\%) & Ref. Frac. RMS (\%) \\
\midrule
 2018-04-18 &  Iqueye at Galileo     &  58226.987512 &  58227.144271 &  3582 &         8694.0 &        8.05 &    3.66 \\\\
 2018-05-21\textsuperscript{*} &  Iqueye at Galileo     &  58259.963819 &  58260.073970 &  1186 &          920.0 &        5.01 &       / \\\\
 2018-06-08 &  Iqueye at Galileo     &  58277.880174 &  58278.072153 &  4483 &         2732.0 &        4.24 &    2.78 \\\\
  \multirow{2}{*}{2018-07-18} &  Iqueye at Galileo     &  58317.857575 &  58318.008258 &  6292 &         3118.0 &        4.89 &    4.39 \\
    &Aqueye+ at Copernicus &  58317.860313 &  58318.079815 &  8080 &        39838.0 &        2.67 &     2.51 \\\\
 2018-07-19 &  Iqueye at Galileo     &  58318.905007 &  58319.067042 &  7192 &         2334.0 &        2.74 &    4.97 \\\\
 2018-07-23 &  Aqueye+ at Copernicus &  58322.888808 &  58323.017569 &  5388 &        35768.0 &        4.79 &     3.78 \\\\
 2018-07-28 &  Iqueye at Galileo     &  58327.868617 &  58327.946255 &  2181 &         4014.0 &        2.67 &     1.6\\\\
 2018-10-13 &  Iqueye at Galileo     &  58404.735278 &  58404.853843 &  5384 &         1125.0 &        4.68 &    3.87 \\
\midrule
\multicolumn{8}{l}{\textsuperscript{*}Bad weather. Reference star not observed.}\\
\end{tabular}
\tablefoot{A typical observing sequence is composed of 1 minute on sky, 15 or 30 minutes on target, and 5 minutes on a reference star. This is repeated several times. The last observation is 1 minute on sky. Since a typical sequence starts and finishes with a sky observation, the start and stop times in the table refer to them rather than to the start/stop time of the observations of the target. However, T$_{obs}$ refers to the time on target, considering only the time with the best sky conditions. The reported count rate is the background-subtracted average rate. In the last column are reported the fractional RMS of a reference star (GSC 00444-02282) in the vicinity of MAXI J1820+070.}
\end{table*}

\begin{table*}[!th]
\centering
\tiny
\caption{Summary of the telescopes involved in the optical photometric observations.}
\label{tab:telescopes}
\begin{tabular}{lll}
    \toprule\toprule
 Telescope ID & Type & Location \\
\midrule
 1400 & 67/92-cm Schmidt telescope & Cima Ekar Observing Station, Asiago (VI), Italy \\
 1301 & 50-cm Ritchey-Chrétien telescope & Osservatorio Astronomico Santa Lucia di Stroncone (TR), Italy \\
 2300 & 40-cm Cassegrain telescope & Stazione Astronomica di Sozzago (NO), Italy \\
 G2   & 60-cm Cassegrain telescope & Stara Lesna Observatory, Slovakia \\
 18   & 18-cm Maksutov telescope & Stara Lesna Observatory, Slovakia \\
 A3   & 50-cm Maksutov telescope & Dibai E.A. Astronomical Station of SAI MSU, Crimea \\
 2T   & 60-cm Cassegrain telescope & Dibai E.A. Astronomical Station of SAI MSU, Crimea \\
 3T   & 125-cm Cassegrain telescope & Dibai E.A. Astronomical Station of SAI MSU, Crimea \\ 
 SA   & 100-cm Cassegrain telescope & Special Astrophysical Observatory of the Russian Academy of Science, Caucasus, Russia \\
\midrule
\end{tabular}
\end{table*}

\begin{table*}[!th]
\centering
\tiny
\caption{Summary of the optical photometric observations.}
\label{tab:log_lowtime}
\begin{tabular}{llllrrrr}
\toprule\toprule
Band & Telescope ID &  Start (MJD) &  Stop (MJD) & Observing Nights & Covered nights (\%) & MAG min &  MAG max \\
\midrule
 \textbf{U} &  A3 &  58270 &  58755 &  13 &  2.7 &  15.580 &  12.628 \\
  &  G2 &  58219 &  58923 &  63 &  8.9 &  16.214 &  11.796 \\
  &  SA &  58240 &  58697 &  14 &  3.1 &  18.438 &  12.087 \\
\midrule
 \textbf{B} &  1301 &  58927 &  59066 &   3 &   2.1 &  19.204 &  15.154 \\
  &  1400 &  58338 &  59066 &  62 &   8.5 &  19.227 &  13.429 \\
  &  2300 &  58295 &  58912 &  75 &  12.1 &  19.107 &  12.728 \\
  &    18 &  58221 &  58738 &  31 &   6.0 &  14.990 &  12.368 \\
  &    2T &  58274 &  58729 &  25 &   5.5 &  15.400 &  13.422 \\
  &    A3 &  58270 &  58766 &  35 &   7.0 &  18.080 &  13.395 \\
  &    G2 &  58219 &  58923 &  67 &   9.5 &  18.841 &  12.362 \\
  &    SA &  58240 &  58740 &  21 &   4.2 &  19.212 &  12.702 \\
\midrule
 \textbf{V} &   1301 &  58927 &  59066 &    7 &   5.0 &  19.183 &  14.802 \\
  &   1400 &  58338 &  59066 &   62 &   8.5 &  18.533 &  13.259 \\
  &   2300 &  58295 &  58912 &   83 &  13.4 &  18.422 &  12.824 \\
  &     18 &  58221 &  58738 &   34 &   6.6 &  14.946 &  12.217 \\
  &     2T &  58274 &  58729 &   26 &   5.7 &  15.184 &  13.234 \\
  &     A3 &  58270 &  58767 &   48 &   9.6 &  17.612 &  13.178 \\
  &  AAVSO &  58198 &  58432 &  139 &  59.1 &  15.337 &  11.806 \\
  &     G2 &  58219 &  58923 &   67 &   9.5 &  16.802 &  12.259 \\
  &     SA &  58240 &  58740 &   21 &   4.2 &  18.219 &  12.577 \\
\midrule
 \textbf{Clear (to V)*} &  AAVSO &  58201 &  58390 &  129 &  67.9 &  14.753 &  11.677 \\
\midrule
 \textbf{R} &  1301 &  58927 &  59066 &   7 &   5.0 &  18.145 &  14.512 \\
  &  2300 &  58295 &  58912 &  84 &  13.6 &  18.073 &  12.641 \\
  &    18 &  58221 &  58738 &  35 &   6.8 &  14.390 &  11.970 \\
  &    2T &  58274 &  58729 &  28 &   6.1 &  14.844 &  12.961 \\
  &    A3 &  58270 &  58766 &  37 &   7.4 &  17.029 &  12.914 \\
  &    G2 &  58219 &  58923 &  72 &  10.2 &  18.469 &  11.874 \\
  &    SA &  58240 &  58740 &  21 &   4.2 &  17.529 &  12.295 \\
\midrule
 \textbf{Clear (to R)*} &  18 &  58580 &  58723 &   2 &    1.4 &  14.317 &  13.557 \\
  &  2T &  58801 &  58801 &   1 &  100.0 &  18.200 &  18.200 \\
  &  3T &  58638 &  58808 &  14 &    8.2 &  18.258 &  16.942 \\
  &  G2 &  58364 &  58923 &  21 &    3.8 &  18.282 &  13.346 \\
\midrule
 \textbf{I} &  18 &  58221 &  58738 &  33 &  6.4 &  14.162 &  11.583 \\
  &  2T &  58274 &  58729 &  26 &  5.7 &  14.430 &  12.604 \\
  &  A3 &  58270 &  58765 &  35 &  7.1 &  16.430 &  12.640 \\
  &  G2 &  58219 &  58923 &  63 &  8.9 &  15.966 &  11.631 \\
  &  SA &  58240 &  58740 &  21 &  4.2 &  16.732 &  11.941 \\
\midrule
 \textbf{g'} &  1400 &  58338 &  59066 &  66 &  9.1 &  18.921 &  13.326 \\
\midrule
 \textbf{r'} &  1400 &  58338 &  59066 &  68 &  9.3 &  18.251 &  13.242 \\
\midrule
 \textbf{i'} &  1400 &  58338 &  59066 &  65 &  8.9 &  18.191 &  13.222 \\
\midrule
\multicolumn{8}{l}{\textsuperscript{*}Clear (to V) or Clear (to R) stands for the observations taken in white light (without any filters) and reduced as if they were taken in V or R band.} \\
\end{tabular}
\tablefoot{The column "Covered nights" shows the percentage of the observed nights over the total number of nights in the interval from the first to the last observation for each band and each Telescope ID (Table \ref{tab:telescopes}).}
\end{table*}

\label{lastpage}
\end{appendix}

\end{document}